\documentclass[aps,fleqn,showpacs,superscriptaddress,amsmath,amssymb,floatfix]{cas-dc}

\usepackage[numbers]{natbib}
\usepackage{microtype}
\usepackage{graphicx}
\usepackage{dcolumn}
\usepackage{bm}
\graphicspath{ {./images/} }
\usepackage{color}
\usepackage{hyperref}
\usepackage{footnote}
\usepackage{wrapfig}
\usepackage{subcaption}
\usepackage{longtable}
\usepackage[T1]{fontenc}
\usepackage{xcolor}

\usepackage{amsmath}
\usepackage{amssymb}
\usepackage{amsthm}
\usepackage{amsfonts}
\usepackage{braket}
\usepackage{tabularx} 
\usepackage{lineno}

\begin{document}
\let\WriteBookmarks\relax
\def\floatpagepagefraction{1}
\def\textpagefraction{.001}
\shorttitle{DyCo$_{5}$}
\shortauthors{Vishina {\it et al}}

\title [mode = title]{New insights into the magnetism of DyCo$_{5}$}                      

\author[1]{Alena Vishina}[orcid=0000-0002-4583-2877]
\cormark[1]
\ead{alena.vishina@physics.uu.se}
\address[1]{Department of Physics and Astronomy, Uppsala University, Box 516, SE-75120, Uppsala, Sweden}

\author[2]{Konstantin Skokov}
\address[2]{Technische Universität Darmstadt, Germany}

\author[3]{Hiroki Tsuchiura}
\address[3]{Department of Applied Physics, Tohoku University, Sendai 980-8579, Japan}

\author[1]{Patrik Thunström}

\author[2]{Alex Aubert}

\author[2]{Oliver Gutfleisch}

\author[1,4]{Olle Eriksson}
\address[4]{WISE-Wallenberg Initiative Materials Science, Uppsala University}

\author[1]{Heike C. Herper}

\cortext[cor1]{Corresponding author}

\begin{abstract}
In this work, we present the first magnetization measurements of DyCo$_5$ single crystals in magnetic fields up to 14 T, spanning a temperature range up to 600 K. Our investigation reveals several unique features, including a significant magnetization anisotropy and an observed minimum in spontaneous magnetization near the compensation point, phenomena not previously reported. This work also uncovers the complex magnetic behavior of DyCo$_5$, with a pronounced interplay between the Dy and Co sublattices, each exhibiting distinct temperature-dependent magnetic properties. The combination of dynamical mean-field theory (DMFT), atomistic spin-dynamics (ASD) simulations, and the Effective Spin Model (ESM) for rare-earth compounds successfully explains the experimental data across both low and high temperatures. Our theoretical approach not only explains the observed magnetic anisotropy and the behavior near the compensation temperature but also successfully reproduces key experimental features such as the saturation behavior at high fields and the evolution of the magnetic moment at different temperatures. 


\end{abstract}




\maketitle

\section{Introduction}

Rare-earth–transition-metal (RE–TM) intermetallic compounds exhibit a rich variety of magnetic properties due to the interplay between the localized 4$f$ moments of rare-earth ions and the itinerant 3$d$ electrons of transition metals \cite{ALGARABEL1988213,1058829,DELMORAL1987285,PhysRevB.101.214433,PhysRevLett.105.137205}. For light RE elements such as Ce, Pr, and Sm, the combination of uniaxial magnetic anisotropy and ferromagnetic coupling between the Co and RE sublattices makes them promising candidates for permanent-magnet applications \cite{REIronPM,Handbook,10.1063/1.2947236,BUSCHOW1974903,MA2015610}. Among these, the 1:5 compound SmCo$_5$ was the first to attract significant attention as a high-performance permanent-magnet material, owing to its strong uniaxial magnetocrystalline anisotropy (MCA) and high Curie temperature \cite{10.1063/1.1657138,10.1063/1.1709459,ma17040808}. Following this discovery, other RECo$_5$-type compounds  were synthesized and studied; however, most did not succeed as permanent magnets due to their insufficient uniaxial or dominant planar magnetic anisotropy \cite{10.1063/1.341084}.

Many members of this family containing heavier rare-earths, including DyCo$_5$ and TbCo$_5$, have attracted considerable interest because of their intriguing temperature-dependent magnetic phenomena, such as spin reorientation transitions and magnetization compensation behavior \cite{TSUSHIMA1983197,BALLOU1989266}. These features arise from the antiferromagnetic exchange coupling between the RE element and the Co sublattice, the competition between the strong planar anisotropy induced by rare-earth ions and the uniaxial anisotropy originating from the Co sublattice, and the contrasting temperature dependence of their respective magnetizations. The preferred magnetization direction of an RE ion is governed by the interplay between the crystal-field (CF) splitting and spin-orbit (SO) interaction. As a result, RECo$_5$ compounds provide a valuable platform for studying complex sublattice interactions in RE–TM systems. 

Spin reorientation transitions offer potential for novel magnetic functionalities, such as thermally driven switching or nonvolatile magnetic control, relevant for magnetic recording technologies \cite{PhysRevApplied.5.064007}, including heat-assisted magnetic recording (HAMR) \cite{PhysRevApplied.5.064007,PhysRevLett.103.117201} and all-optical magnetic recording \cite{PhysRevLett.99.047601}. Additionally, these materials hold promise for applications in laser-induced ultrafast demagnetization \cite{doi:10.1142/S2010324715500046}, thermally induced magnetization switching at the magnetization compensation point \cite{10.1038}, spintronics, and other emerging technologies \cite{spintronics,doi:10.1142/S2010324717400100,PhysRevB.95.180409}. Such applications motivate a renewed investigation of DyCo$_5$ using modern experimental and theoretical techniques.

One of the earliest studies on the magnetic behavior of DyCo$_5$ was conducted by Tsushima and Ohokoshi in the early 1980s~\cite{TSUSHIMA1983197}. Using single-crystalline samples, they identified both a magnetization compensation point ($T_{\mathrm{comp}}$) and a spin reorientation transition ($T_{\mathrm{SR}}$), providing strong evidence of complex magnetic interplay between the Dy and Co sublattices. Their results laid the foundation for subsequent investigations of RECo$_5$ compounds exhibiting similar behavior. However, due to the limitations of experimental techniques at the time, the resolution, angular selectivity, and quantitative reliability of the magnetization data were restricted, making detailed comparisons with theoretical models difficult.

It is noteworthy that all previous measurements of the magnetization of DyCo$_5$ single crystals were conducted in magnetic fields not exceeding 2 T \cite{1059178,OHKOSHI1977195,TSUSHIMA1983197}. While such fields are sufficient to reveal magnetization anomalies associated with the spin-reorientation transition, they are considerably lower than the anisotropy field of DyCo$_5$. As a result, these earlier studies are inadequate for reliably evaluating the respective contributions of magnetocrystalline anisotropy and intersublattice exchange interactions, thereby limiting a comprehensive understanding of the underlying physics of the transition. 

In recent years, both experimental and theoretical advances have enabled revisiting such classical systems from a modern perspective (see further details in {\it Appendix} C). In particular, high-precision magnetization measurements on well-characterized single crystals can now resolve anisotropic components of the magnetization with sufficient accuracy to capture subtle temperature-dependent features. At the same time, theoretical frameworks based on first-principles calculations and atomistic modelling have matured to the point where they can provide quantitative predictions for finite-temperature magnetic properties. In this context, Donges \textit{et al.}~\cite{PhysRevB.96.024412} reported a multiscale modelling study of DyCo$_5$ thin films, combining XMCD measurements and first-principles-based simulations. While their results provide valuable insight, our present work focuses on bulk single crystals, which reveal unique features such as the absence of full compensation and a significant variation in magnetization with the direction of the applied magnetic field, offering a complementary view of the intrinsic magnetic behavior of DyCo$_5$.

 A significant challenge in computational studies of RE–TM materials lies in the treatment of the strongly correlated 4$f$ shell, since conventional electronic-structure approaches such as the Local Density Approximation (LDA) or the Generalized Gradient Approximation (GGA) often fail to describe correlated-electron systems accurately. Several methods have been developed to address this issue, including DFT+$U$ \cite{Anisimov_1997,PhysRevB.69.134408,PhysRevB.94.014303}, the “open-core” scheme in which 4$f$ electrons are removed from the valence band \cite{Brooks_1991,BUTCHER20171,PhysRevB.96.100404,10.1063/1.4996989}, the self-interaction correction (SIC) method \cite{PhysRevLett.120.097202,PhysRevB.97.224415,PhysRevB.71.205109}, and dynamical mean-field theory (DMFT) \cite{Galler,RevModPhys.78.865,PhysRevB.106.224411}. 

In this work, we present a comprehensive investigation of the finite-temperature magnetic properties of DyCo$_5$, combining precise magnetization measurements on single-crystalline samples with two complementary theoretical approaches. The experimental data, obtained with explicit separation of the $c$-axis and the in-plane components, provide a detailed view of the magnetization behavior, including the compensation point, spin reorientation transition, and the evolution of magnetic anisotropy. Studies performed on DyCo$_5$ bulk single crystals offer distinct advantages over thin-film investigations of magnetic anisotropy, owing to their well-defined crystallographic orientation, structural homogeneity, similar demagnetization factor along all directions, and the absence of substrate-induced effects. In such crystals, the intrinsic magnetocrystalline anisotropy can be probed directly, free from complications associated with strain, interfacial diffusion, or texture that typically affect thin films. Furthermore, bulk specimens allow exploration over a broader temperature range -- up to $\sim$1000 K -- enabling a detailed examination of anisotropy evolution up to the Curie temperature.

On the theoretical side, we combine two distinct modelling strategies. The first is a hybrid approach, regarded as an effective spin model (hereafter referred to as ESM), which incorporates a microscopic crystal-field theory for Dy$^{3+}$ ions based on parameters obtained from first-principles calculations, together with a phenomenological description of the Co-sublattice magnetization using the empirical formula proposed by Kuz’min \cite{10.1063/1.4927849}. This framework captures the key mechanisms underlying spin reorientation and anisotropy evolution with relatively simple input parameters. The second approach combines density functional theory (DFT), dynamical mean-field theory (DMFT), atomistic spin dynamics (ASD), and input from the ESM, providing a first-principles-based, multiscale simulation framework that allows direct computation of temperature-dependent magnetic properties. By employing and comparing these theoretical results with experimental observations, we assess the capabilities and limitations of each approach and explore their complementarity in describing the magnetic behavior of DyCo$_5$.

An overview of previous experimental and theoretical work on DyCo$_5$ is provided in {\it Appendix} \ref{sec:app3}.

The structure of this paper is as follows. {\it Section}~\Ref{sec:exp_method} details the synthesis, structural characterization, and magnetization measurements performed on DyCo$_{5}$ single crystals. {\it Section}~\Ref{sec:th_method} outlines the computational approaches employed, including the DFT+DMFT+ASD multiscale framework and the effective spin model for rare-earth-transition-metal systems. {\it Section}~\Ref{sec:main} presents a comprehensive comparison between experimental results and theoretical predictions, focusing on the evolution of magnetization, magnetocrystalline anisotropy, and spin reorientation behavior. Finally, {\it Section} \ref{sec:disc} summarizes the main outcomes of this study and critically evaluates the performance and limitations of the theoretical models in describing the magnetic properties of DyCo$_5$.

\section{Experimental methods}\label{sec:exp_method}

DyCo$_5$ ingots were prepared by induction melting under an Ar atmosphere. The ingots were then sealed in a quartz tube filled with Ar and annealed at 1050$^{\circ}$C for 16 days, followed by quenching in water. Single crystals several millimeters in size were subsequently extracted from the annealed ingots. The phase purity of the grains was confirmed using backscatter electron (BSE) imaging and energy-dispersive X-ray spectroscopy (EDS) performed with a {\it Tescan Vega 3} scanning electron microscope (SEM) equipped with an {\it EDAX Octane Plus} detector. 

To facilitate the extraction of single-crystalline grains from the ingot, a weak aqueous solution of acetic acid was used to etch the grain boundaries. Surface defects were removed by electrochemical polishing in a 70\% H$_3$PO$_4$ solution at room temperature, with an applied voltage in the range of 6-12 V. The best DyCo$_5$ single crystal was selected for magnetic measurements and oriented by back-scattering Laue X-ray diffraction along the {\it c}-axis (sixfold symmetry axis [001]) and the {\it a}([100]) and {\it b'} ([120])  crystallographic directions.

Magnetic measurements were carried out using a 14 T physical properties measurement system ({\it PPMS, Quantum Design}) equipped with a vibrating sample magnetometer. Measurements were performed under varying temperatures and magnetic fields, with the magnetic field swept at a rate of 0.01 T/s.

\section{Computational methods}\label{sec:th_method}

The complex and often unconventional properties of materials composed of RE and transition metal (TM) elements arise from the indirect coupling between the localized RE-4$f$ and TM-3$d$ magnetic moments, mediated via the RE-5$d$ orbitals. While the transition metal sublattice provides high magnetization, the localized 4{\it f} electrons are responsible for high magnetocrystalline anisotropy \cite{perm_mag,6008648,SKOMSKI2009675}. To account for the complex magnetic interactions in DyCo$_5$ and to achieve accurate reproduction of the experimental results, we employed two types of calculations. In the first, we used the ESM for the RE sublattice to describe the magnetic behavior, incorporating experimental data as input. The second approach combines DMFT and ASD, which does not rely on experimental input other than the crystal structure. However, it benefits from additional input from the ESM, particularly the temperature dependence of the magnetocrystalline anisotropy.

\subsection{DFT+DMFT}

For the DFT and DMFT calculations, we employed the full-potential linear muffin-tin orbital (FP-LMTO) method as implemented in the {\it RSPt} code \cite{rspt1,rspt2}, using the PBE functional \cite{PBE} for exchange and correlation. Brillouin-zone integrations were performed using the Fermi smearing method. We used an extension of the LDA+DMFT implementation \cite{PhysRevB.76.035107,PhysRevB.79.165104} based on FP-LMTO, which includes full self-consistency in the electronic density, as described in detail in Refs.~\cite{GRANAS2012295,PhysRevB.94.085137}.

The Hubbard-I approximation (HIA) \cite{PhysRevB.57.6884,Hubbard,SVANE2006364} was used as the impurity solver for the Dy 4$f$ orbitals, while the fully renormalized spin-polarized T-matrix fluctuation-exchange (SPTF) solver was employed to treat the Co 3$d$ states, following its successful application to SmCo$_5$ in Ref.~\cite{GRANAS2012295}. The required Slater parameters for the Co 3$d$ orbitals were also adopted from that work.

Experimental unit-cell parameters were used together with a $20\times20\times20$ $k$-point mesh. DyCo$_5$ crystallizes in the hexagonal CaCu$_5$-type structure, with one Dy atom occupying the 1$a$ site and five Co atoms located at the 2$c$ and 3$g$ sites.



For the spin-dynamics calculations, we used Metropolis Monte Carlo (MC) and Atomistic Spin Dynamics (ASD) simulations as implemented within the Uppsala  Atomistic  Spin  Dynamics (UppASD) software \cite{ASD}. The simulations were performed on a $25\times 25\times 25$ supercell with periodic boundary conditions. 

The spin Hamiltonian used in UppASD included the following terms: Heisenberg exchange, uniaxial magnetic anisotropy, and the external magnetic field, expressed as

\begin{equation}\label{eq:HASD}
{\mathcal H}=-\sum_{ij} J_{ij}\mathbf{e}_i\cdot\mathbf{e}_j-\sum_{i} K_{1}(\mathbf{e}_i\cdot\mathbf{e}^K_i)^2-\sum_{i} B^{ext}\mathbf{e}_i, 
\end{equation}
where $i$ and $j$ denote atomic indices, $\mathbf{e}_i=\frac{\mathbf{m}_i}{m_i}$ is the unit vector of the atomic magnetic moment $\mathbf{m}i$, $J_{ij}$ is the strength of the exchange interaction, $K_1$ is the uniaxial anisotropy constant along the direction $\mathbf{e}^K_i$, $B^{ext}$ is the external field.

The exchange parameters for the ASD simulations were calculated with the {\it RSPt} code for the first five coordination shells \cite{PhysRevB.91.125133, Liechtenstein_1984,Heisenberg} using the Lichtenstein-
Katsnelson-Antropov-Gubanov (LKAG) method \cite{LIECHTENSTEIN198765}. 
Since SO coupling in DyCo$_5$ cannot be neglected, the relativistic effects were taken into account to compute the components of the intersite magnetic interactions. The full derivation and implementation are described in Ref.~\cite{PhysRevB.102.115162}. For instance, if the magnetic moment points along the {\it z}-axis, the {\it xx}-component can be written as

\begin{equation}\label{eq:HASD}
J_{ij}^{xx}=\frac{T}{4}\sum_{n}\text{Tr}_{m}[\widehat{\mathcal{H}}_{i},\widehat{\sigma}^{y}]G_{ij}(i\omega_{n})[\widehat{\mathcal{H}}_{j},\widehat{\sigma}^{y}]G_{ji}(i\omega_{n}), 
\end{equation}
where $\widehat{\mathcal{H}}$ is the electronic Hamiltonian, $G_{ji}$ are Green's functions, $\widehat{\sigma}$ is the vector of Pauli matrices, and $\omega_{n}$ are Matsubara frequencies. The trace is taken over orbital indices denoted by $m$, and $T$ represents temperature. 

In the commutators, the spin-orbit coupling term in the Hamiltonian $\widehat{\mathcal{H}}$ induces a change not only in the spin orientation but also in the orbital momentum of the electrons. This causes the electrons to change their orbitals which affects the chemical bonding. To avoid this, we have implemented a filtering step where only the spin exchange terms are kept in the commutator. A comparison of magnetization curves with and without this filtering is provided in {\it Appendix} E. In the main results, however, the 'filtered' exchange constants are employed.

The temperature-dependent anisotropy constants $K_1(T)$ obtained from the ESM were used as input for the ASD simulations.

In addition, the {\it Sumo} package \cite{sumo} and the {\it VESTA} code \cite{vesta} were utilized for visualisation. 

\subsection{ESM for RE Intermetallics}


The distinction between the roles of the 3$d$ transition-metal (TM) electrons and the 4$f$ electrons of rare-earth (RE) elements in determining magnetism and magnetocrystalline anisotropy (MCA) has long been a fundamental concept in RE–TM compounds.

To capture this division of roles for DyCo$_5$, we employ a two-sublattice phenomenological Hamiltonian~\cite{yamada1988}
\begin{equation}
    {\mathcal H}
    = \sum_{i} \hat{\mathcal H}_{{\rm eff},i}
    + K_{1}^{\rm TM}(T) \sin^{2}\theta_{\rm TM}
    - {\bm M}^{\rm TM}(T) \cdot {\bm B} ,
\label{eq:H_total}
\end{equation}
where the first term is the RE single-ion contribution, the second represents the uniaxial anisotropy of the TM sublattice, and the third is the Zeeman energy of the TM magnetization ${\bm M}^{\rm TM}$. Here $\theta_{\rm TM}$ denotes the angle between ${\bm M}^{\rm TM}$ and the crystallographic $c$ axis, and in the present system the easy axis is along $c$.

For each RE site $i$, we diagonalize
\begin{align}
\hat{\mathcal H}_{{\rm eff},i} &=
\lambda\, \hat{\bm L}_{i} \cdot \hat{\bm S}_{i}
+ 2\mu_{\rm B}\, \hat{\bm S}_{i} \cdot {\bm B}_{{\rm ex},i}(T)
+ \sum_{l,m} B_{l,i}^{m}\, \hat O_{l,i}^{m}
\nonumber \\
&~~ + \mu_{\rm B}\,(\hat{\bm L}_{i}+2\hat{\bm S}_{i}) \cdot {\bm B} ,
\end{align}
restricted to the ground-state multiplet $J=15/2$ of Dy$^{3+}$. In this expression, $\hat{\bm L}_{i}$ and $\hat{\bm S}_{i}$ denote the orbital and spin angular-momentum operators of the Dy$^{3+}$ 4$f$ shell, and $\mu_{\rm B}$ is the Bohr magneton. The crystal-field term uses the Stevens operator-equivalent form $\sum_{l,m} B_{l,i}^{m}\hat O_{l,i}^{m}$, where $\hat O_{l,i}^{m}$ are Stevens operators and $B_{l,i}^{m}$ are the Stevens crystal-field parameters in the $J$-manifold. The Stevens parameters relate to the well-known crystal-field coefficients as
\[
B_{l,i}^{m} = \bigl(A_{l,i}^{m}\langle r^{l}\rangle\bigr)~\theta_{l}(J),
\]
with $\theta_{l}(J)$ the Stevens factors for ranks $l=2,4,6$, denoted $\alpha_{J},\beta_{J},\gamma_{J}$ for $J=15/2$ \cite{Stevens1952,Hutchings1964}.

The vector ${\bm B}_{{\rm ex},i}(T)$ denotes the TM-induced exchange field acting on RE site $i$. Within a standard mean-field description it is taken collinear with the TM sublattice magnetization and independent of $i$ in magnitude, so that ${\bm B}_{{\rm ex},i}(T)=B_{\rm ex}(T)\,{\bm e}_{\rm TM}$ with $B_{\rm ex}(T)\equiv \lvert{\bm B}_{{\rm ex},i}(T)\rvert$. We further assume a linear scaling of its magnitude with the TM magnetization,
\[
B_{\rm ex}(T)=\bigl[B_{\rm ex}(0)/M_{\rm TM}(0)\bigr]\,M_{\rm TM}(T),
\]
which implies
\[
\frac{M_{\rm TM}(T)}{M_{\rm TM}(0)}=\frac{B_{\rm ex}(T)}{B_{\rm ex}(0)}\equiv \alpha(T).
\]

For a given temperature $T$ and TM easy-axis direction ${\bm e}_{\rm TM}$, the single-site partition function and free-energy contribution are defined by
\[
Z_{i}=\sum_{n} e^{-E_{n,i}/k_{\rm B}T}, ~~~
g_{i}(T,{\bm e}_{\rm TM})=-k_{\rm B}T\ln Z_{i}.
\]

Expanding $g_{i}$ for a small tilt of ${\bm e}_{\rm TM}$ from the $c$ axis,
\[
g_{i}(\theta)=g_{i}^{(0)}+\tfrac12 g_{i}^{(2)}\theta^{2}+\tfrac{1}{4!} g_{i}^{(4)}\theta^{4}+\cdots,
\]
we identify the RE single-ion contributions to the uniaxial anisotropy energy as
\[
k_{1,i}(T)=\tfrac12 g_{i}^{(2)}(T), ~~
k_{2,i}(T)=\tfrac{1}{4!} g_{i}^{(4)}(T)+\tfrac13 k_{1,i}(T).
\]
The macroscopic constants per formula unit are
\[
K_{1}(T)=\frac{1}{V_{0}}\sum_{i} k_{1,i}(T)+K_{1}^{\rm TM}(T), ~~
K_{2}(T)=\frac{1}{V_{0}}\sum_{i} k_{2,i}(T),
\]
where $V_{0}$ is the volume per formula unit.

For the TM sublattice we use the Kuz'min formula for the reduced magnetization,
\[
\alpha(T)=\Bigl[
1 - s\!\left(\frac{T}{T_{\rm C}}\right)^{3/2}
  - (1-s)\!\left(\frac{T}{T_{\rm C}}\right)^{p}
\Bigr]^{1/3},
\]
where $T_{\rm C}$ is taken from experiment for DyCo$_5$.
We set $s=0.7$ and $p=5/2$ as in the Kuz'min parametrization for YCo$_{5}$~\cite{Kuzmin2005}, a member of  the same $R$Co$_{5}$ family as DyCo$_{5}$.
For the TM contribution to the anisotropy we adopt the Callen–Callen single-ion scaling \cite{CALLEN1960310} for leading uniaxial $l=2$ term,
\[
\frac{K_{1}^{\rm TM}(T)}{K_{1}^{\rm TM}(0)}
=\left[\frac{M_{\rm TM}(T)}{M_{\rm TM}(0)}\right]^{3}
=\alpha^{3}(T),
\]
which we use as a minimal approximation.

\label{eq:kuzmin}




\section{Results}\label{sec:main}

\subsection{Experimental results}\label{sec:exp_results}

\begin{figure*}[h]
 \centering
 \includegraphics[scale=0.68]{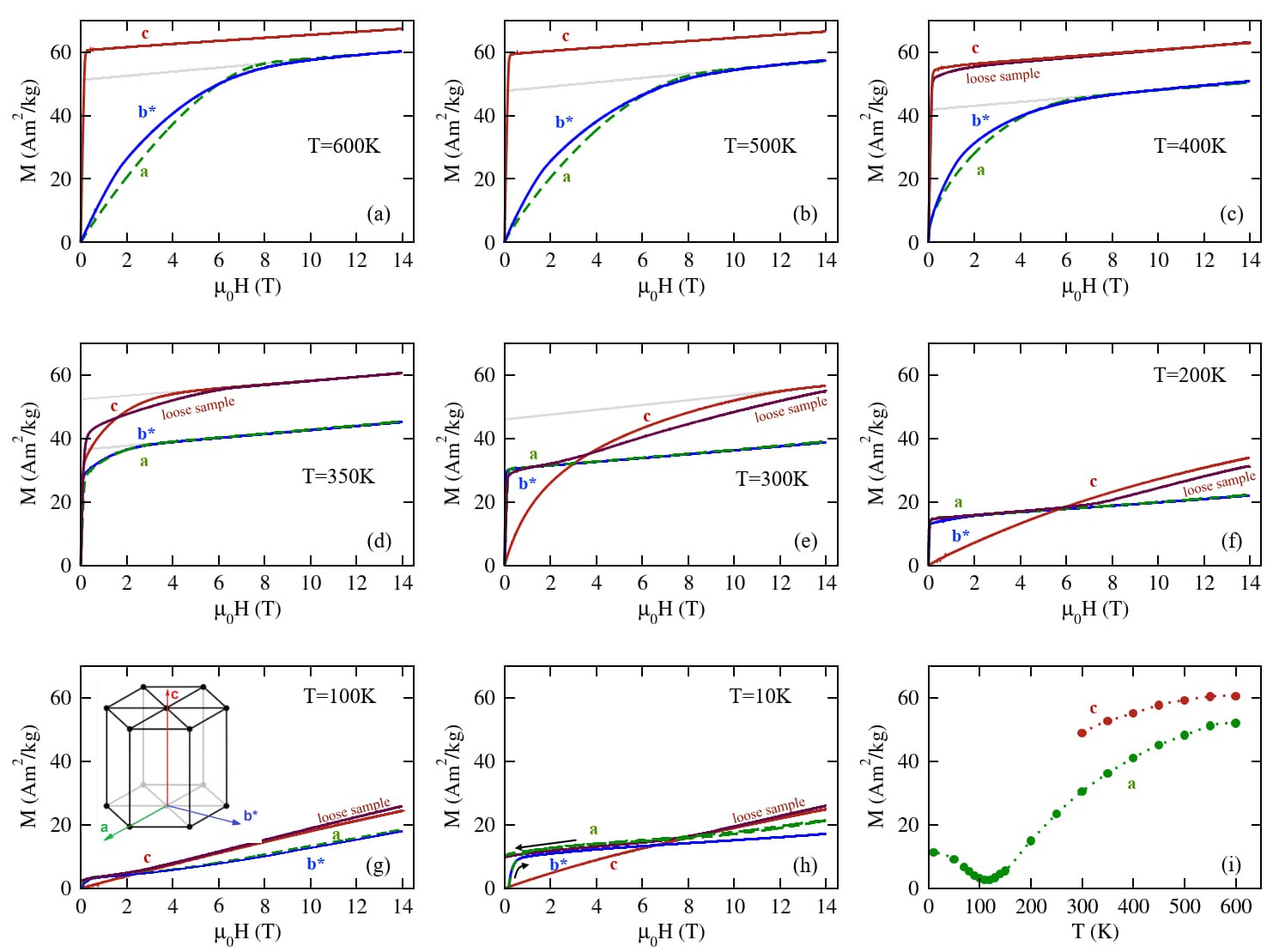}
\caption{(Color online) {\it (a-h)} Magnetisation curves of DyCo$_{5}$ single crystal measured along the three principal crystallographic directions in magnetic fields up to 14 T and within a temperature range of 10-600K. Magnetisation curves of a loose sample are also shown in {\it (c-h)}. {\it (i)} Spontaneous magnetization measured along the {\it a} and {\it c} directions (extrapolated to the demagnetisation field)}
\label{Exp_res}
\end{figure*}

Fig.~\ref{Exp_res} {\it (a-h)} shows magnetization curves measured along the three principal crystallographic directions in magnetic fields up to 14 T, within a temperature range of 10-600 K. For $T=$ 400 K and higher {\it (a-c)}, DyCo$_{5}$ exhibits uniaxial magnetic anisotropy, with the anisotropy field varying from approximately 6 T at 400K to 8 T at 600K. In systems with strong magnetic anisotropy, the anisotropy field is defined as the field above which the sample reaches magnetic saturation when magnetized along the hard axis \cite{10.1063/1.4927849,BOLYACHKIN2017740}. 

The data reveal significant magnetization anisotropy intrinsic to this compound: the magnetization along the hard direction does not reach the same value as along the easy axis - even beyond the anisotropy field - so that a finite gap between the two curves persists up to the highest fields in Fig.~\ref{Exp_res} {\it (a-c)}. Spontaneous magnetization was deduced from the easy axis magnetization curves by extrapolation to the demagnetizing field, whereas magnetization along the hard axis in a low magnetic field was obtained by extrapolating the magnetisation along the {\it a}-axis above the anisotropy field to the demagnetizing field. These extrapolations are shown as dotted lines in Fig.~\ref{Exp_res} {\it (a–h)}.  At 400K, the spontaneous magnetization along the {\it c}-axis is 55.2 Am$^{2}$/kg, whereas along the {\it a}-axis it is 41.07 Am$^{2}$/kg. 

This behavior is primarily attributed to two factors. First, the thermal agitation of atomic magnetic moments varies between the easy axis (EA) and hard axis (HA) directions \cite{CALLEN1960310}. Second, the degree of quenching of the orbital angular momentum of the 3{\it d} atoms varies across crystallographic directions \cite{rozenfel1995low}. This phenomenon is particularly pronounced in high-anisotropy intermetallic compounds composed of rare-earth elements and 3{\it d} transition metals  \cite{ALAMEDA19801257,BOLYACHKIN2017740,10.1063/1.4936604,10.1063/1.4927849,KARPENKOV20181012}. Significant magnetization anisotropy has been observed in RECo$_{5}$ compounds such as YCo$_{5}$ (4\%) and CeCo$_{5}$ (10-12\%). In comparison, the DyCo$_{5}$ exhibits exceptionally high magnetization anisotropy: at 300K, the magnetization anisotropy reaches 31\% and decreases with temperature to 8.5\% at 600K.

The type of magnetic anisotropy in DyCo$_{5}$ changes between 370 K and 325 K. Upon cooling, the ferrimagnetic moment deviates from the {\it c}-axis, passing through an easy-cone magnetic anisotropy regime. The spin-reorientation transition starts at 367 K (uniaxial anisotropy $\rightarrow$ easy-cone) and completes at 325 K (easy-cone $\rightarrow$ easy-plane anisotropy). As seen in Fig.~\ref{Exp_res} {\it (e,f)}, above the crossing point of magnetisation curves measured along the {\it c} and {\it a}-axes, the magnetic moment along {\it c} remains significantly higher than along {\it a}, indicating strong anisotropy of the magnetic moment below the spin reorientation temperature. 

In addition to the measurements on a fixed sample, magnetization curves of a freely rotating sample were recorded between 400 K and 10 K. At 400 K (Fig.~\ref{Exp_res} {\it (c)}), the magnetization curve of the loose sample nearly coincides with that measured along the easy axis. However, as the temperature decreases, the field dependence becomes more complex. In the low-field region, the magnetization follows the easy-axis ({\it a}-axis) behavior, but above the anisotropy field, the sample rotates relative to the {\it c}-axis because the magnetic moment along {\it c} exceeds that along {\it a}. This response indicates that in strong fields, the magnetization of the freely rotating sample is not aligned along the easy axis -- an unusual effect that reflects the giant magnetization anisotropy of DyCo$_{5}$.

At 110 K, the sample passes through a compensation point where the magnetizations of the Co and Dy sublattices become equal due to their different temperature dependences. Notably, complete antiferromagnetic compensation between the Dy and Co sublattices was not observed: the spontaneous magnetization reaches a minimum value of 2.5 Am$^{2}$/kg at 110 K and increases again upon further cooling. At 100 K (Fig.~\ref{Exp_res} {\it (g)}), 10 K below the compensation point, the magnetization curves in high fields show a noticeable slope, indicating a non-collinear magnetic structure. The magnetic moments of the Co and Dy sublattices are not strictly antiparallel but form an angle smaller than 180$^{\circ}$, a characteristic feature of ferrimagnets near the compensation point. At 10 K (Fig.~\ref{Exp_res} {\it (h)}), the non-collinearity of the sublattices --     manifested as an increase in magnetization in all directions -- is again observed above 10 T, and a small magnetic hysteresis is also present at this temperature.

\subsection{Magnetism - Effective Spin Model}\label{sec:hir_results}

\begin{figure}[htbp]
  \begin{center}
    \includegraphics[width=85mm]{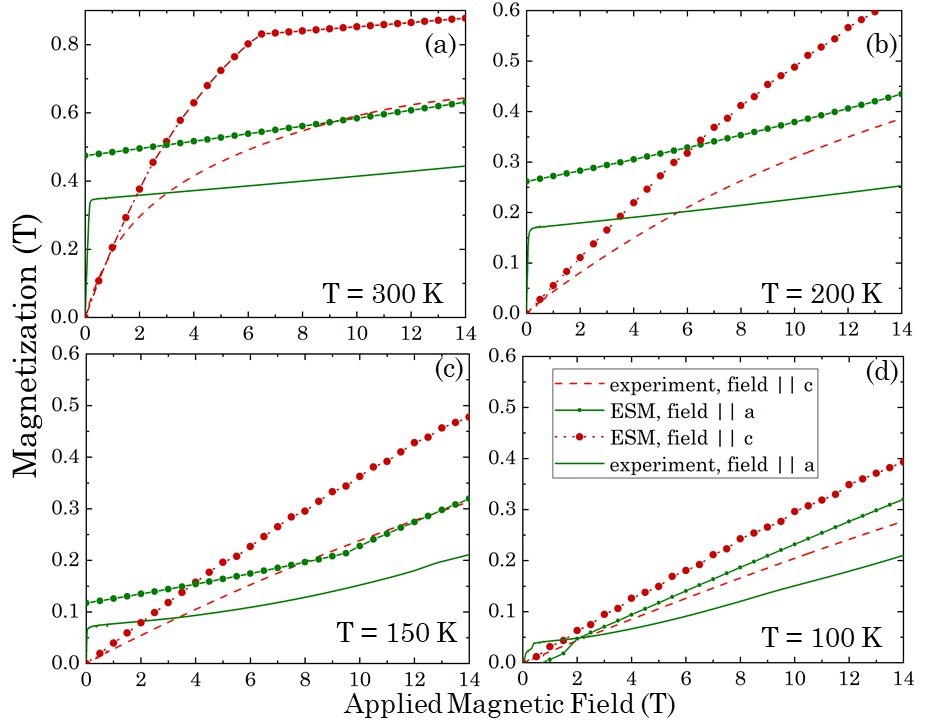}
    \caption{Magnetization curves of DyCo$_{5}$ at 100 K, 150 K, 200 K, and 300 K calculated within the framework of the effective spin model and compared with the experimental data. The green and red lines indicate results along the principal [100] and [001] crystallographic directions, respectively.} \label{fig:mag_curve}
  \end{center}
\end{figure}

The experimental trend of the magnetization as a function of the applied magnetic field is well reproduced by the ESM. Figure~\ref{fig:mag_curve} (a)--(d) shows the calculated magnetization curves for four representative temperatures, spanning from the compensation temperature to the spin-reorientation transition. For comparison, the corresponding experimental data are also plotted in each panel.

The temperature dependence of the spontaneous magnetization, $M(T)$, in zero applied field is discussed in more detail later (see Fig.~\ref{M_T} (a)), in connection with the magnetization compensation behavior and the spin-reorientation transition of DyCo$_{5}$.


In addition, the calculated temperature dependence of the magnetocrystalline anisotropy constants $K_{1}$ for Co and Dy sublatices is shown in  Fig.~\ref{Ktheory}.
The Co sublattice in the RECo$_{5}$ structure exhibits a moderate uniaxial anisotropy along the $c$-axis, as indicated by the fact that $K_{1}^{\mathrm{Co}}$ remains positive over the entire temperature range. However, its magnitude is significantly smaller than the strong in-plane anisotropy introduced by Dy ions at low temperatures. This behavior is readily understood from the shape of the 4$f$ electron cloud determined by the Stevens factor. The list of calculated crystal field parameters is provided in Table~\ref{table:ani}, along with the values reported in previous studies.

As demonstrated above, the effective Hamiltonian provides a convenient and reasonably accurate framework for describing the finite-temperature magnetic properties of DyCo$_{5}$. It captures the essential features of the magnetization behavior, offers a satisfactory level of quantitative agreement, and shows the dominance of Dy $K_{1}$ at low temperatures. However, the results in Fig.~\ref{fig:mag_curve} clearly indicate that further improvement is required, particularly in incorporating the thermal fluctuations of the 3$d$ electron system in the Co sublattice.

One step in this direction is to use the ESM output in the ASD framework.

\begin{table*}[h!]
\centering
\caption{Magnetocrystalline anisotropy and crystalline field parameters data reported in previous investigations. 'CW' stands for 'current work'.}
\begin{tabular}{c c c c c c c c} 
 \hline \hline
K$_{1}^{Dy}$ &  K$_{1}^{Co}$ & A$_{2}^{0}$ & A$_{4}^{0}$ & A$_{6}^{0}$ & A$_{6}^{6}$ & B$_{2}^{0}$ & Ref. \\ 
mRy & mRy & mRy & mRy & mRy & mRy &  \\
 \hline
 & 0.075 (2c), 0.04 (3g) & & & & & & YCo$_5$ \cite{FRANSE1993307,etde_6658043} (exp) \\ 
 & 0.06 & -1.6 & -0.01 & & 0.11 & & \cite{FRANSE1993307,etde_5447691}  \\ 
 -0.97 (0 K) & & & & & & 0.006 & \cite{RADWANSKI1986120} \\ 
 & & -4.7 & & & & 0.0057 & \cite{GREEDAN1973387}  \\ 
 & & -3.62 & -0.12 & 0.016 & -0.38 &  & CW (ESM)  \\ 
\hline \hline
\end{tabular}
\label{table:ani}
\end{table*}

\subsection{DFT+DMFT+ASD description of magnetism }\label{sec:upp_res}

To gain deeper insight into the sublattice magnetism and to account for temperature fluctuations and external magnetic field effects, we employed a combination of DFT, DMFT, and ASD simulations. The temperature dependence of the element-resolved $K_{1}$, obtained with the ESM, was incorporated into the spin dynamics simulations.

\subsubsection{DFT+DMFT}

%
The DMFT calculations were performed, treating the correlated and localized Dy 4{\it f}-electrons within the Hubbard-I Approximation \cite{PhysRevB.57.6884, Hubbard, SVANE2006364}. 
Assuming an in-plane orientation of the spin polarization axis -- consistent with the experimentally observed easy-plane anisotropy at low temperatures -- the total magnetic moment of the system was calculated to be 1.54 $\mu_{\bf B}$/f.u (at 47 K). The Dy atom carries a total moment of 9.81 $\mu_{\bf B}$, comprising a 4$f$ spin moment of 4.85~$\mu_{\mathrm{B}}$, a 4$f$ orbital moment of 4.77~$\mu_{\mathrm{B}}$, and a 5$d$ magnetic moment of 0.20~$\mu_{\mathrm{B}}$. For the Co sublattices, the moments are $m_{\mathrm{Co(I)}} = -1.79~\mu_{\mathrm{B}}$ and $m_{\mathrm{Co(II)}} = -1.73~\mu_{\mathrm{B}}$ (47 K). These values are in good agreement with experimental data and are summarized in Table~\ref{table:overview}, together with the previously reported results (A more detailed discussion of earlier experimental and theoretical studies on DyCo$_{5}$ is provided in {\it Appendix}~\ref{sec:app3}). 

As the material undergoes a spin reorientation, we also performed calculations assuming a uniaxial (out-of-plane) orientation of the magnetic moments. This change significantly affects the Dy moment, which decreases to 8.30 $\mu_{\bf B}$ (comprising 4.35~$\mu_{\mathrm{B}}$ from the 4$f$ spin moment, 3.77~$\mu_{\mathrm{B}}$ from the 4$f$ orbital moment, and 0.18~$\mu_{\mathrm{B}}$ from the 5$d$ magnetic moment). For the Co sublattice, the corresponding values are $m_{Co(I)}$ = -1.80 $\mu_{\bf B}$ and $m_{Co(II)}$ = -1.76 $\mu_{\bf B}$. These results are also listed in Table~\ref{table:overview}.

\begin{table*}[h!]
\centering
\caption{Magnetization (total and sublattice-resolved), magnetocrystalline anisotropy, compensation temperature, reorientation temperatures, and Curie temperatures for the previous and the current work. '(0 K)' denotes extrapolation to 0 K. The first Co moment corresponds to the 2{\it c} site, while the second one gives the moment for the 3{\it g} site. 'CW' stands for 'current work'.}
\begin{tabular}{l c c c c c c c c c c} 
 \hline \hline
M$_{tot}$ &  M$_{Dy}^{tot}$ &  M$_{Dy}^{spin}$ & M$_{Dy}^{orb}$ & M$_{Co}^{orb}$ & M$_{Co}^{tot}$ & T$_{\bf comp}$ & T$_{\bf SR1}$ & T$_{\bf SR2}$ & T$_{\bf C}$& Ref. \\ 
$\mu_{\bf B}$/f.u. & $\mu_{\bf B}$ & $\mu_{\bf B}$ & $\mu_{\bf B}$ & $\mu_{\bf B}$ & $\mu_{\bf B}$ & K & K & K & K  & \\
 \hline
1.64 (0 K) & & & & & & 147 &  &  &  & \cite{10.1063/1.1728809} (exp) \\ 
1.7 (0 K) & 8.3 (150 K) & & & & & 170 &  &  &  & \cite{PhysRev.140.A131} (exp) \\ 
1.40 (0 K) & & & & & & 148 & 325 & 367 & 925 & \cite{TSUSHIMA1983197} (exp) \\ 
 & 8.8 & & & & -1.77, -1.72 & & & & & \cite{DyCo5exp} (exp) \\
 & 9.5 (0 K) & 4.47 & 4.97 & -0.13 & -1.75 & 120 & 320 & 360 & & \cite{PhysRevB.96.024412} (exp) \\ 
1.1 & & & & & & 123 & & & 960 & \cite{Buschow_1977} (exp) \\
1.00 & & & & & & 135 & & & & \cite{1059178} (exp) \\ 
1.76 & 9.5 & &  & & -1.58, -1.50   & 164 & 432 & 460 & 1030 & \cite{PhysRevB.96.024412} (sim) \\ 
 & 8.65 & 4.77 & 3.84 & & -1.52, -1.48 & & & & & \cite{MILETIC2007604} (sim) \\ 
 & 9.77 & &  & & -1.61, -1.50   &  &  &  &  & \cite{SMARDZ1999209} (sim) \\ 
 1.52 & 10.26 & 5.33 & 4.93 & -0.26 & -1.49 (av)   & 85  &  &  & 870 & \cite{PhysRevB.97.224415} (sim) \\ 
1.12 & 9.81 & 5.07 & 4.74 & -0.19, -0.13 & -1.79, -1.73 & 150 & 235 & 240 & 754 & CW (DMFT+ASD) \\ 
 & 8.30 & 4.35 & 3.77 & -0.18 & -1.76 &  &  &  &  & CW (DMFT M$_S\perp$c) \\ 
1.36 & 9.85 & & & & 1.70 & 110 & 356 & 374 & 1030 & CW (ESM) \\ 
0.92 & & & & & & 110 & 325 & 367 & & CW (exp) \\ 
\hline \hline
\end{tabular}
\label{table:overview}
\end{table*}

To investigate magnetic-moment behavior at higher temperatures and under external magnetic fields, ASD simulations were carried out using relativistic $J_{ij}$ (for details, see Section \ref{sec:th_method}). Two sets of $J_{ij}$ values were calculated -- corresponding to in-plane and out-of-plane orientations of the magnetic moments -- and both were used to obtain the magnetization curves shown in Fig.~\ref{M_T_a_c}.



\begin{figure}[h]
 \centering
 \includegraphics[scale=0.42]{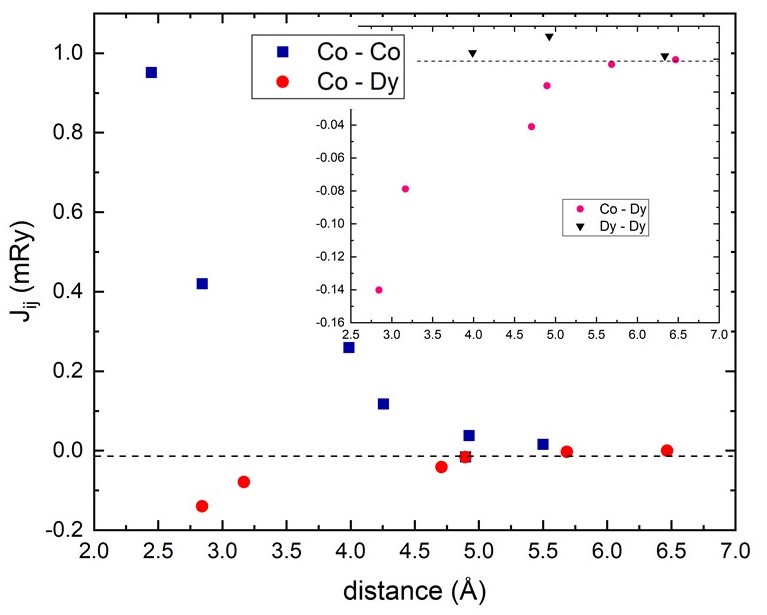}
\caption{(Color online) Exchange interactions $J_{ij}$ in DyCo$_{5}$ calculated with RSPt. The inset shows $J_{ij}$ for Dy-Co and Dy-Dy interactions.}
\label{fig:Jij}
\end{figure}

As expected, the Co atoms are coupled ferromagnetically, whereas the interactions between Co and Dy are antiferromagnetic, as shown in  Fig.~\ref{fig:Jij}. The Co-Co interactions are approximately an order of magnitude stronger than the Co-Dy interactions, while the Dy-Dy coupling is negligible. The calculated exchange parameters for the nearest neighbors (NN) are $J_{NN}^{Co-Co}=0.95$ mRy, $J_{NN}^{Co-Dy}=-0.140$ mRy, $J_{NN}^{Dy-Dy}=0.004$ mRy. The $J_{NN}^{Co-Co}$ value agrees well with \cite{PhysRevB.96.024412}, while our $J_{NN}^{Co-Dy}$ values are somewhat smaller. It should be noted, however, that slight disorder was used in \cite{PhysRevB.96.024412} to fit the experimental data. Our calculated T$_{\bf C}$ underestimates the experimental value. As discussed in {\it Appendix}~\ref{sec:app1}, this discrepancy arises primarily from the absence of explicit temperature dependence in the exchange parameters $J_{ij}$.

\subsubsection{ASD: Temperature effects}

The output of the DMFT+HIA calculations, namely, the exchange parameters $J_{ij}$ and magnetic moments, was used as input for the ASD calculations (see {\it Computational methods} for details). In this section, we examine the evolution of the magnetic state of DyCo$_{5}$ with temperature and applied magnetic field to gain further insight into our experimental observations.

\begin{figure}[h]
 \centering
 \includegraphics[scale=0.35]{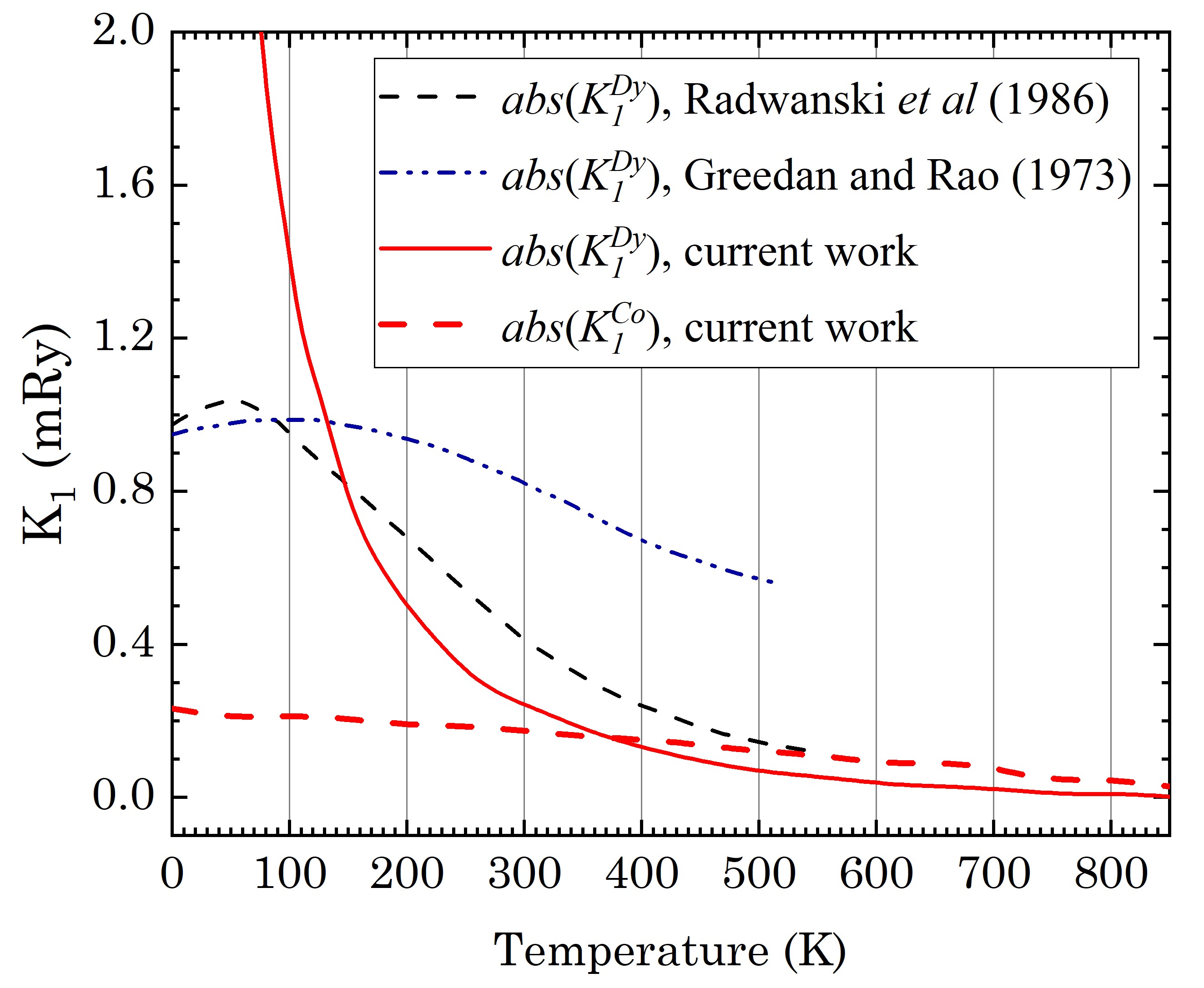}
\caption{(Color online) Absolute values of $K_{1}$ for Co and Dy as a function of temperature calculated in this work in the ESM. Magnetocrystalline anisotropy calculated within the single-ion approximation by Radwanski in \cite{RADWANSKI1986120} as well as by Greedan and Rao \cite{GREEDAN1973387} are shown as well. Note, that Dy shows in-plane (negative values) anisotropy, unlike Co.}
\label{Ktheory}
\end{figure}

\begin{figure}[!tb]
 \centering
 \includegraphics[scale=0.54]{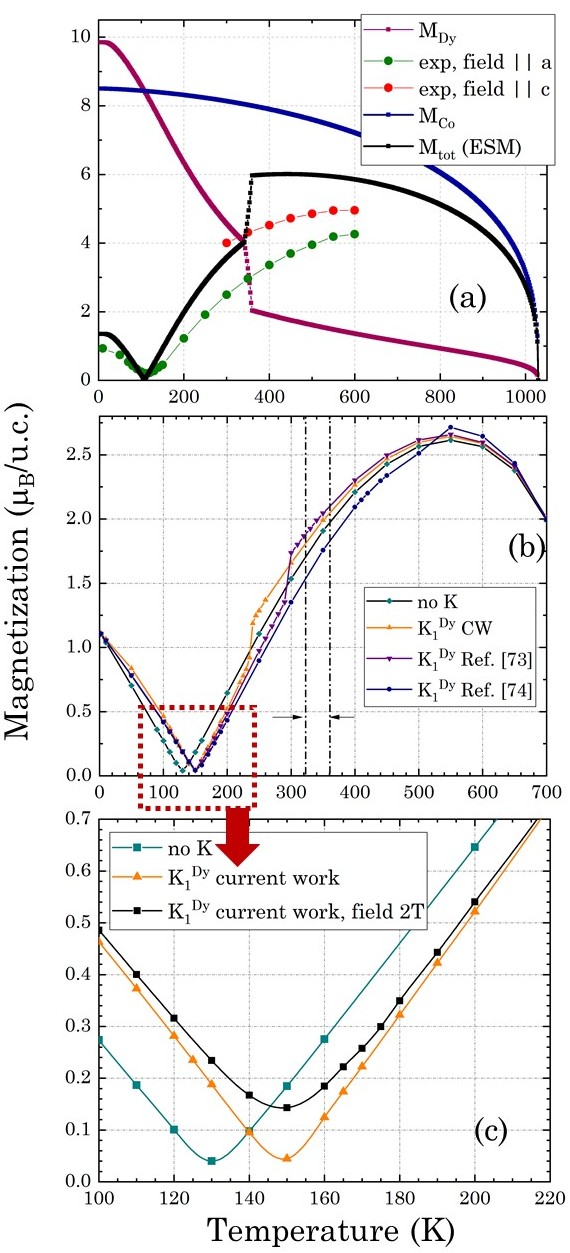}
\caption{(Color online) (a) Temperature dependence of the magnetic moments of Co and Dy, as well as the total moment, obtained by using the effective spin model. (b) Temperature dependence of the magnetic moment per unit cell obtained with ASD simulation for three different calculated $K_{1}(T)$. The vertical dotted lines show the experimental interval between T$_{\bf SR1}$ and T$_{\bf SR2}$. (c) Temperature dependence of the magnetic moment per unit cell around T$_{\bf comp}$ with and without magnetic anisotropy ($K_{1}(T)$ calculated in the current work). For the former case, {\it M(T)} is also shown with and without the external field of 2 T. Note that there is residual magnetization around the compensation point. The Hamiltonian used in ASD simulations is shown as Eq.~\ref{eq:HASD}. } 
\label{M_T}
\end{figure}

\begin{figure*}[!tb]
 \centering
 \includegraphics[scale=0.55]{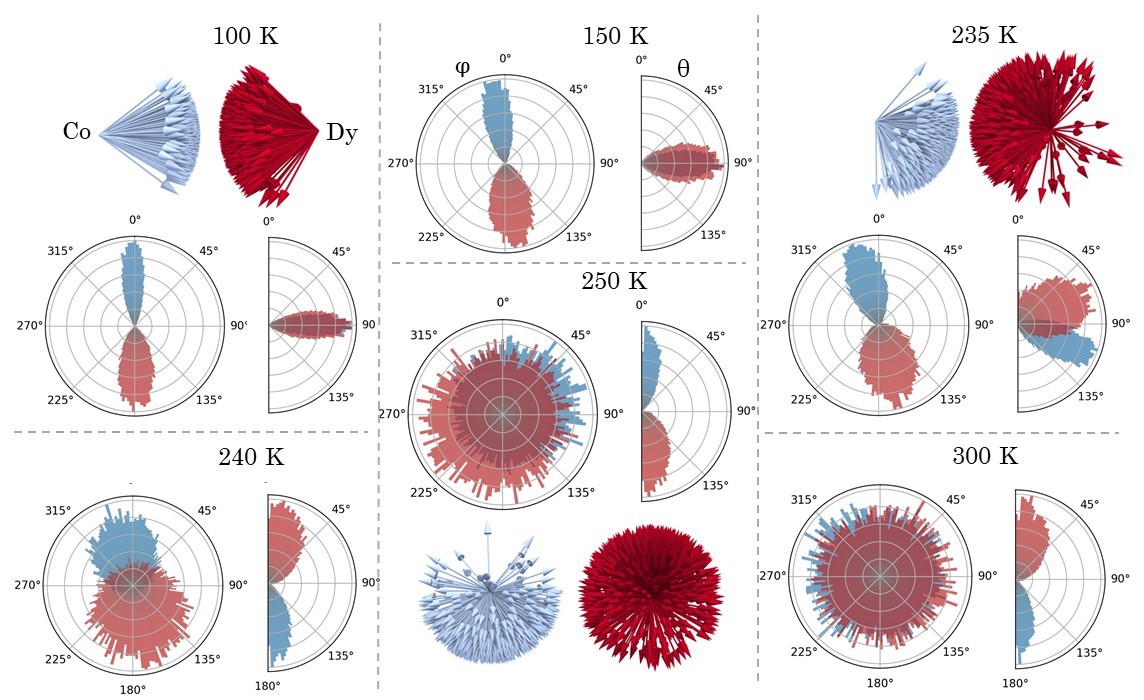}
\caption{(Color online) Spatial distribution of the magnetic moments at different temperatures calculated with $K_{1}(T)$ obtained in the current investigation (with ESM, see Fig.~\ref{Ktheory}). In the diagrams given in spherical coordinates, the length of the vectors represents the number of spins pointing in a chosen direction. Hence, as 5 times more abundant, the longer vectors belong to Co. The color intensity corresponds to the size of the magnetization. 3D versions of the same distribution are given for visual clarity for 100K, 225 K, and 250 K. Here, magnetization vectors are normalized to 1, blue vectors show Co moments, while red arrows correspond to Dy. For the images, 4\% of the ASD supercell vectors are shown with the common origin.}
\label{M_T_polar}
\end{figure*}

The temperature-dependent anisotropy constants $K_{1}(T)$ for Co and Dy obtained in the present work (from the ESM) were employed, along with those calculated within the single-ion approximation in Refs.~\cite{RADWANSKI1986120,GREEDAN1973387}, for comparison. All three $K_{1}(T)$ functions are shown in Fig.~\ref{Ktheory}, and the corresponding temperature dependences of magnetization -- together with the result obtained without magnetocrystalline anisotropy -- are plotted in Fig.~\ref{M_T} (b). The overall trends are similar for all cases; however, distinct differences emerge near the compensation point and the SRT. Introducing anisotropy increases the compensation temperature in all three cases, from approximately 130~K to 150~K. This effect can be understood from the microscopic origin of the compensation point. The compensation arises from the different thermal responses of the two sublattices: the Dy sublattice is strongly affected by thermal fluctuations and exhibits a wide angular distribution of moments, whereas the Co sublattice remains relatively well ordered (see the visualization in Fig.~\ref{M_T_polar}). The strong anisotropy term constrains the Dy moments to the plane, counteracting thermal fluctuations and thus slowing the reduction of the Dy sublattice moment.
In contrast, the reorientation interval depends sensitively on the choice of $K_{1}(T)$. Using the $K_{1}(T)$ obtained in this work, we underestimate the reorientation temperature, predicting a transition between 235 K and 240 K, compared to the experimental range of 320-367 K \cite{PhysRevB.96.024412,TSUSHIMA1983197}. The $K_{1}(T)$ calculated in \cite{GREEDAN1973387} leads to an overestimation of the reorientation temperatures, while T$_{\bf SR1}$ and T$_{\bf SR2}$ calculated using $K_{1}(T)$ from \cite{RADWANSKI1986120} are slightly below the experimental range.


Fig.~\ref{M_T} (c) presents an enlarged view of the region around the compensation temperature for the magnetization curves calculated with and without magnetocrystalline anisotropy $K_{1}(T)$. For comparison, an additional set of data is shown for the magnetization in an external field of 2 T applied along the {\it y}-axis.  As observed experimentally, the magnetization does not drop completely to zero near the compensation temperature, and the residual magnetization increases with the applied field.

\begin{figure}[!tb]
 \centering
 \includegraphics[scale=0.38]{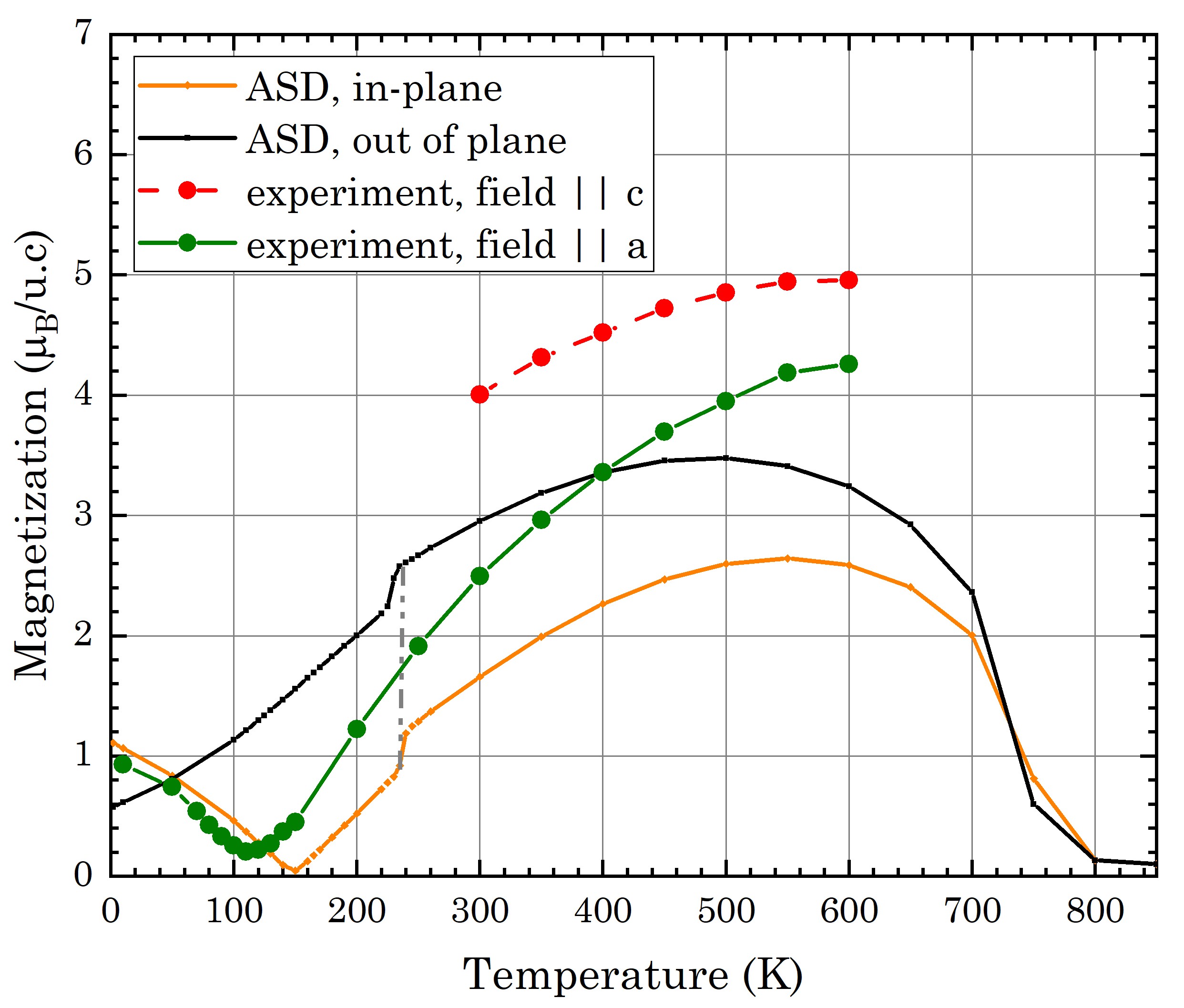}
\caption{(Color online) Magnetization of the unit cell of DyCo$_{5}$ as a function of temperature obtained experimentally (for two orientations of the magnetic field) and with ASD simulations. Computational results are shown with the two sets of inputs from DMFT calculations where the spin axis was set in-plane and along the {\it c}-axis (magnetic moments and $J_{ij}$). K$_1$ obtained in the current work was used in both simulations. The dotted line shows the change in the moment as magnetization rotates from in-plane to out-of-plane.} 
\label{M_T_a_c}
\end{figure}

A similar behaviour is obtained when using the magnetic  moments and exchange parameters J$_{ij}$ calculated for the uniaxial orientation of the spins (see Fig.~\ref{M_T_a_c}). In this configuration, with the spin orientation parallel to the {\it c}-axis, the magnetization remains significantly higher after the spin-reorientation region, consistent with our experimental observations.

The spatial distribution of the Dy and Co magnetic moments at various temperatures provide additional insight into the evolution of the total magnetization, particularly around the compensation point. The corresponding diagrams, calculated in the absence of the external magnetic field, are shown in Fig.~\ref{M_T_polar}. As reported in the previous studies, below the compensation temperature (e.g. 100 K) the magnetic moments lie in the {\it ab}-plane, since the in-plane magnetocrystalline anisotropy of Dy exceeds the combined uniaxial anisotropy of five Co atoms. Owing to the AFM exchange interactions, the Dy and Co moments are oriented antiparallel to each other. As the temperature increases, the rare-earth moment decreases much more rapidly than the {\it 3d}-element moment, and the total magnetization reaches its minimum around 150 K, where the opposing sublattice contributions nearly cancel each other. At appoximately 235 K, the magnetization axis begins to rotate toward the {\it c}-axis, signaling the onset of the spin-reorientation process. At higher temperatures, the uniaxial anisotropy becomes dominant, and the magnetization aligns fully along the {\it c}-axis.

\subsubsection{External field effect}

To directly compare the experimental and theoretical results, we performed ASD simulations under applied external magnetic fields. Experimentally, DyCo${5}$ does not reach magnetic saturation even in fields up to 12 T. In the simulation, we examined fields applied both along the {\it c}-axis and perpendicular to it (within the {\it ab}-plane) at 50 K and 400 K. For these simulations, we used the magnetic moments and exchange intergals obtained from the DMFT+HIA for the in-plane orientation of the spin-axis, together with the temperature-dependent anisotropy constants K$_1$(T) from the ESM.

\begin{figure*}[!tb]
 \centering
 \includegraphics[scale=0.55]{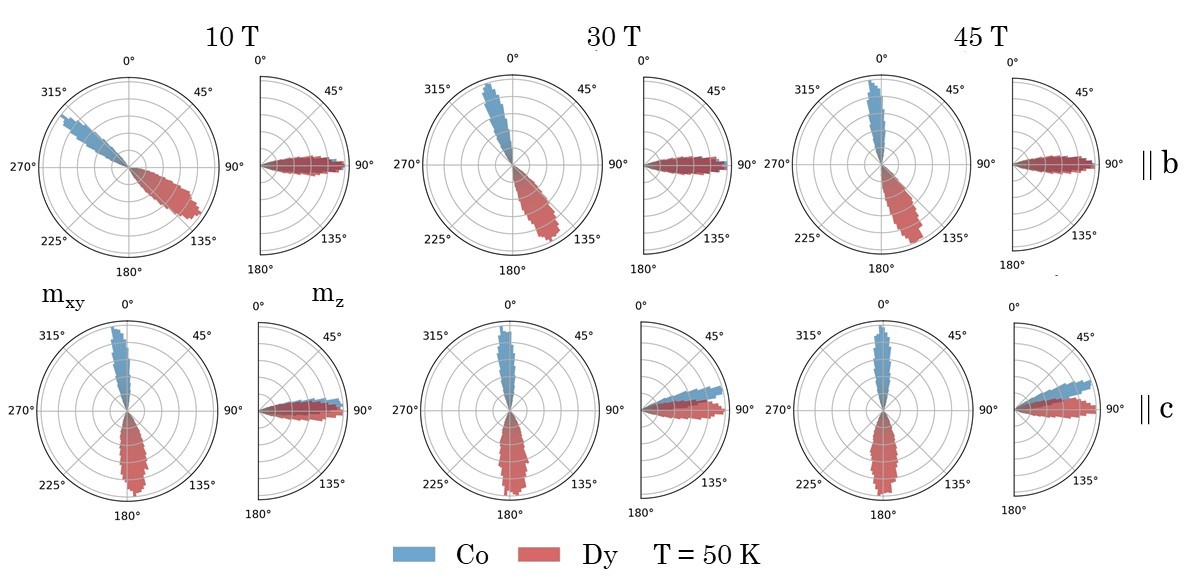}
\caption{(Color online) Spatial distribution of the magnetic moments of Dy and Co in an external field of 10T, 30T, and 45T, with the field oriented in-plane along {\it y} (top) and out-of-plane along {\it z} (bottom) axes. The temperature is 50K, K$_1$(T) used was obtained with ESM (Fig.~\ref{Ktheory}). }
\label{MvsH50}
\end{figure*}

When the magnetic field is applied within the plane (top row of Fig.~\ref{MvsH50}, with the field direction $\phi = 90$\textdegree, $\theta = 0$\textdegree), the Dy and Co moments initially align at an angle relative to the field, pointing in opposite directions. Only at higher fields -- above approximately 25 T -- do the Dy and Co moments begin to noticeably cant towards each other. Even when the applied field is increased to 45 T, the two sublattices remain far from being collinear.

For the field applied along the {\it c}-axis (bottom row of Fig.~\ref{MvsH50}), the Dy and Co moments gradually rotate toward the field direction, still maintaining an antiparallel alignment in the plane. The Co moments begin to cant earlier than Dy, owing to their significantly smaller in-plane anisotropy. In both field orientations, the system fails to reach complete magnetic saturation within the experimental field range, and this behavior persists even at much higher simulated fields.

\begin{figure*}[!tb]
 \centering
 \includegraphics[scale=0.55]{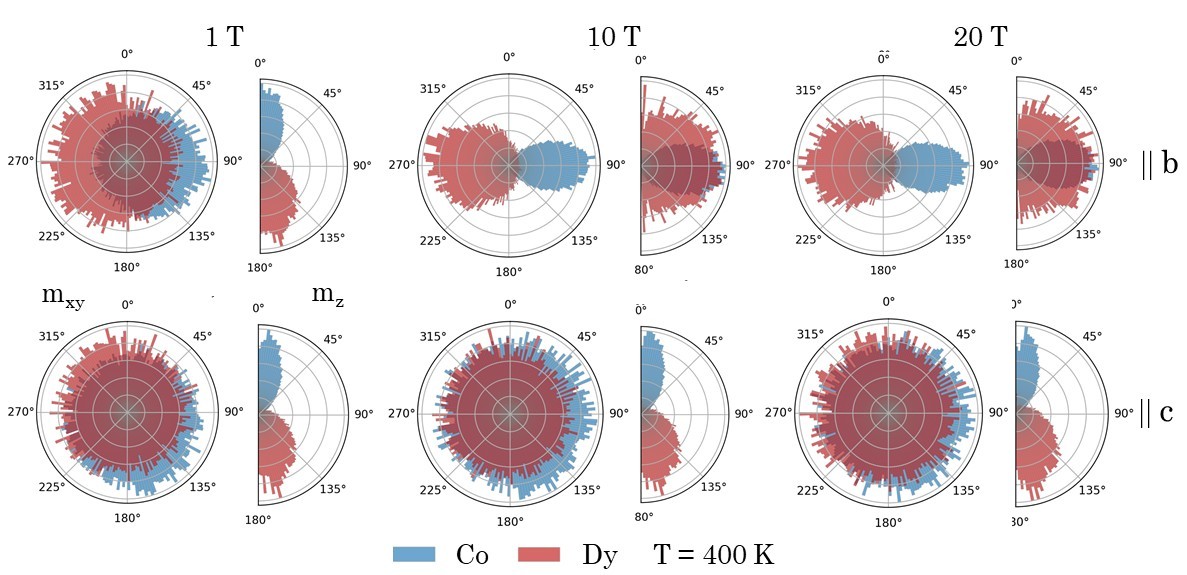}
\caption{(Color online) Spatial distribution of the magnetic moments of Dy and Co in an external field of 10T, 30T, and 45T, with the field oriented in-plane along {\it y} (top) and out-of-plane along {\it z} (bottom) axes. The temperature is 400K. }
\label{MvsH390}
\end{figure*}

We also investigated the magnetic behavior of the system above the spin-reorientation temperature. The spatial distribution of the magnetic moments for several field strengths at $T=400$ K is shown in Fig.~\ref{MvsH390}. In this case, the magnetic moments and exchange integrals obtained from DMFT+HIA for the out-of-plane spin orientation were used. At higher temperatures, the magnetocrystalline anisotropy is substantially reduced, which makes it possible to achieve magnetic saturation for both field orientations.

\section{Discussion and conclusions}\label{sec:disc}

This work presents the first magnetization measurements of DyCo$_{5}$ single crystals performed in magnetic fields up to 14 T across a temperature range up to 600 K. Previous measurements had been limited to fields not exceeding 2 T. High-quality single crystals are essential for resolving the intrinsic magnetic anisotropy in rare-earth-3{\it d} intermetallic compounds. Their well-defined crystallographic orientation enables precise quantification of directional magnetic properties and the respective 4{\it f} and 3{\it d} sublattice contributions. Such measurements are indispensable for disentangling the interplay between crystal electric field effects, spin-orbit coupling, and exchange interactions (including intersublattice exchange), thereby revealing the microscopic origin of magnetic anisotropy and the governing magnetocrystalline energy landscape.

Above 300 K, the spontaneous magnetization along the {\it c}-axis is clearly higher than along the {\it a}-axis. This difference persists even as the easy magnetization direction switches from {\it a} to {\it c} at T$_{\bf SR}\sim$ 350 K, demonstrating the presence of pronounced magnetization anisotropy in this compound. Moreover,  the unique combination of {\it (1)} the compensation point at 110 K, {\it (2)}  significant magnetization anisotropy, and {\it (3)} substantial magnetocrystalline anisotropy complicates the accurate experimental determination of the magnetocrystalline anisotropy of DyCo$_{5}$ below room temperature. Another novel observation -- previously unpredicted by theory -- is the occurrence of a minimum in the spontaneous magnetization near T$_{\bf com}$, without complete antiferromagnetic compensation. To elucidate these features, we employed first-principles-based calculations.

The complex magnetic behavior of DyCo$_{5}$ originates from the distinct temperature dependencies of the Dy and Co sublattices. At low temperatures, both the magnetic moment and magnetocrystalline anisotropy of Dy dominate, whereas at higher temperatures the Co sublattice becomes increasingly influential, being less susceptible to thermal fluctuations. This interplay first leads to magnetic-moment compensation and subsequently to the reorientation of the magnetization axis. Before analyzing the temperature evolution of the sublattice and total magnetizations within the DMFT+ASD framework, it is crucial to determine the temperature dependence of the anisotropy constants, as the magnetocrystalline anisotropy directly enters the ASD Hamiltonian. To this end, we employed the effective spin model (ESM), a well-established approach for estimating magnetocrystalline anisotropy based on crystal field theory combined with 'open-core' first-principles calculations \cite{Richter_1998}, was be applied \cite{Hiroki_18,Hiroki_18_2,Hiroki_21}. 
The resulting magnetization and K$_1$(T) curves for DyCo$_{5}$ obtained with the ESM are shown in  Fig.~\ref{M_T} and  Fig.~\ref{Ktheory}. As expected, at low temperatures, K$_1^{Dy}$(T) is considerably larger than that of the Co sublattice. However, it decreases sharply with increasing temperature and falls below K$_1^{Co}$(T) within the spin-reorientation temperature window.

We find that the $M_{ESM}$(T) curve reproduces the experimental temperature dependence of magnetization well, slightly overestimating both the total moment and the Curie temperature. The in-field magnetization curves of DyCo$_{5}$ calculated using the ESM are shown in Fig.~\ref{fig:mag_curve}. 
The qualitative behavior - such as the saturation fields and the evolution of field dependence for the [001] and [100] directions at different temperatures - captured accurately. However, the theoretical curves overestimate the magnetization by approximately 40\% across most temperature ranges. This discrepancy is not unexpected, as the ESM is essentially an extension of the single-ion model. Overcoming this limitation remains one of the key theoretical challenges for future work.

To go beyond the ESM and gain a deeper insight into the sublattice behavior under varying temperatures and magnetic fields, we perform ASD simulations. These calculations allowed us to trace the evolution of the spin orientations and the distribution of Dy and Co moments. The temperature-dependent anisotropy constant K$_1$ obtained from the ESM was used as input for the ASD simulations.


The DMFT calculations, where the correlated and localised 4{\it f}-electrons were treated within the Hubbard-I Approximation  \cite{PhysRevB.57.6884,Hubbard,SVANE2006364}, have previously demonstrated excellent accuracy for strongly correlated {\it f}-electron systems such as SmCo$_5$ \cite{GRANAS2012295}, REFe$_{12}$, Nd$_{2}$Fe$_{12}$B \cite{PhysRevMaterials.6.084410},  and REFe$_{12}$X hard magnets (RE = Nd, Sm and X = N, Li) \cite{PhysRevB.96.155132}, and Ce-based '1–12' systems \cite{Galler}. The same approach was therefore applied to the experimental unit cell of DyCo$_{5}$. The calculated zero-temperature magnetic moment,  particularly the Co sublattice moment, agrees well with the experiment. Moreover, the calculations successfully reproduce the experimentally observed change in total magnetic moment as the spin axis rotates from in-plane to the out-of-plane ($c$-axis) direction. The difference in the value of magnetic moment is attributed to the effect of the quadrupole moment, for details see {\it  Appendix}~\ref{sec:app2}.

In the ASD simulations, we initialized the system in a FM configuration, and therefore the low-field magnetization process observed experimentally cannot be directly reproduced. Capturing this regime would require micromagnetic simulations, which lie beyond the present scope. Nevertheless, based on the magnetization behaviour produced with ASD, we can explain the differences observed for the two field orientations at different temperatures. 

Figures~\ref{MvsH50} and \ref{MvsH390}help clarify the experimental observations shown in Fig.~\ref{Exp_res}. At low temperatures (see Fig.~\ref{MvsH50}, bottom row), when the field is applied along the {\it c}-axis, the large anisotropy constant prevents the Dy moments from rotating out of the plane, even at high fields (our simulations extend up to 45 T, far exceeding the experimental range). As the field increases, the Co sublattice gradually rotates toward the field direction, leaving the Dy moments behind. Consequently, the total moment increases steadily as the sublattice magnetizations deviate from perfect antiparallel alignment. At higher temperatures (see the 400 K in Fig.~\ref{MvsH50}), the anisotropy becomes uniaxial, K$_{1}^{Dy}$ decreases, and the magnetization aligns along the easy {\it c}-axis, making full saturation possible.


The behavior differs for an in-plane field: at 50 K (below the reorientation temperature; Fig.~\ref{MvsH50}, top row), the moments initially lie within the {\it ab}-plane at an angle to the external field, with Dy and Co antiparallel -- consistent with experimental data below 0.6 T. Further increase in magnetic field leads to the bending of $M_{Dy}$ and $M_{Co}$ toward each other, leading to a nonlinear rise in the total magnetization. 

For fields along the {\it c}-axis, there is no preferred in-plane orientation, and this may vary between grains in the sample. Consequently, no 'initial' low-field increase is observed at low temperatures.


At 400 K (for the FM configuration), all moments align along the easy {\it c}-axis. When the field is applied along {\it c} (Fig.~\ref{MvsH390}, bottom row), the system magnetizes readily, consistent with experiment. In contrast, for an in-plane field (top row of Fig.~\ref{MvsH390}), much stronger fields are required to rotate the spins away from their preferred orientation. This agrees closely with the experimental data in Fig.~\ref{Exp_res}, {\it (a)-(c)}.

As mentioned previously, DMFT calculations were performed for two spin-axis orientations (in-plane and out-of-plane), which significantly affected the Dy magnetic moment. 
The ASD simulations utilized inputs from both orientations, and the results are presented in Fig.~\ref{M_T_a_c} alongside the experimental data. The in-plane-based simulation describes the low-temperature regime below the spin-reorientation point, where the magnetization cannot be rotated out of the plane by experimentally accessible fields. At higher temperatures, experimental points exist for both in-plane and out-of-plane saturation. The simulated magnetization curves successfully reproduce the anisotropy of magnetization observed experimentally.
The remaining quantitative discrepancies at higher temperatures (also reported in Ref.~\cite{PhysRevB.96.024412}), as well as the underestimated Curie temperature, are attributed to the change in the absolute values of the magnetic moments and the exchange constants. For details, see {\it  Appendix}~\ref{sec:app1}.

Our spin-dynamics simulations (see Fig.~\ref{M_T}) reproduce the experimentally observed magnetization minimum at the compensation point. As shown in Fig.~\ref{M_T_polar}, near the compensation temperature the Dy sublattice already exhibits a broad angular distribution of moments, while the Co sublattice remains comparatively well aligned. Moreover, a slight tilt is observed between the Co and Dy sublattices across the full temperature range, see {\it  Appendix}~\ref{sec:app4}. These effects become even more pronounced under external magnetic fields, as shown in Fig.~\ref{M_T}, and prevent perfect compensation of the opposing sublattice moments.

In summary, we have performed high-field magnetization measurements on DyCo${5}$ single crystals, revealing several previously unobserved effects, including pronounced magnetization anisotropy and an incomplete compensation of sublattice moments. Our multiscale theoretical approach -- combining DMFT, ASD, and the effective spin model -- successfully reproduces and explains the key experimental features, providing a comprehensive understanding of the complex magnetic behavior of DyCo$_5$.

\section{Acknowledgement}

The authors would like to acknowledge the support of the Swedish Foundation for Strategic Research, the Swedish Energy Agency, the Swedish Research Council, The Knut and Alice Wallenberg Foundation, eSSENCE, StandUp, and the ERC (synergy grant FASTCORR, project 854843), MaMMoS (EU HORIZON, project no. 101135546). A. Vishina acknowledges the support of Scandinavia-Japan Sasakawa Foundation.

This work was supported by the Deutsche Forschungsgemeinschaft (DFG, German Research Foundation) under Project 405553726-TRR270 and the German Federal Ministry of Education and Research (BMBF)
under Grant BMBF-Projekt05K2022.

In addition, the computations in this work were partially performed using the facilities of the National Academic Infrastructure for Supercomputing in Sweden (NAISS) (projects NAISS2024-5-75, NAISS2024-5-427, and NAISS2024-1-18). 

\section{Competing interests}

The authors declare no competing interests.

 \bibliographystyle{elsarticle-num} 
\bibliography{cas-refs}

\onecolumn
\appendix

\newpage

\section{Magnetization absolute value and J$_{ij}$ change with temperature as the cause of the discrepancies in the ASD simulations}\label{sec:app1}

Figure~\ref{M_T_a_c} shows that the ASD simulations reproduce the overall temperature dependence of magnetization well, but a quantitative mismatch remains in the total moment values -- particularly at higher temperatures -- as well as in the Curie temperature. While the low-temperature results agree closely with the experimental data, we sought to identify the source of the discrepancies observed at elevated temperatures.

\begin{figure}[h]
 \centering
 \includegraphics[scale=0.7]{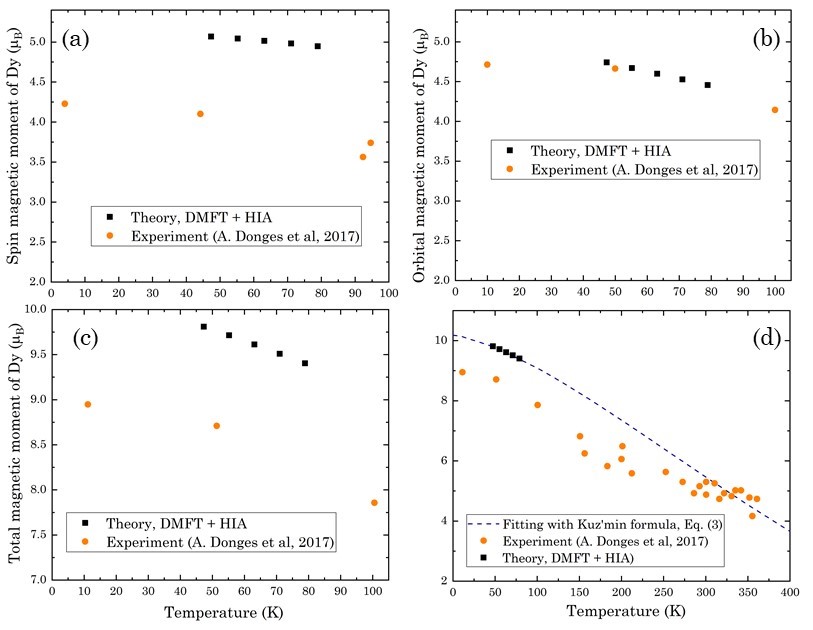}
\caption{(Color online) Temperature dependence of Dy spin (a), orbital (b), and total (c-d) moments calculated with DMFT+HIA. Experimental data are taken from \cite{PhysRevB.96.024412}. In (d), the DMFT data are fitted using the Kuz'min formula, Eq.~\ref{eq:kuzmin}. } 
\label{Dymom}
\end{figure}

\begin{figure}[h]
 \centering
 \includegraphics[scale=0.6]{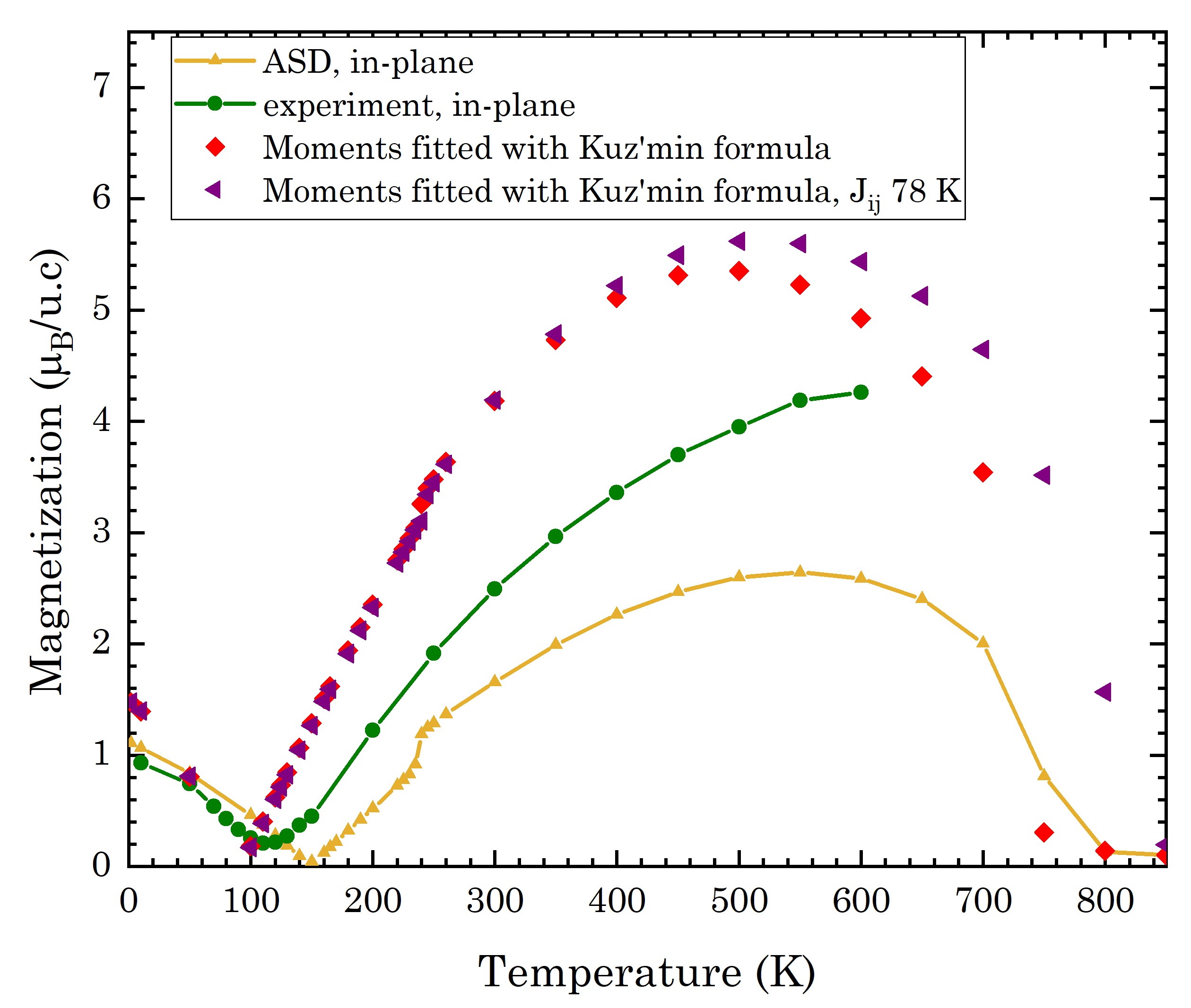}
\caption{(Color online) Temperature dependence of the unit cell magnetic moment of DyCo$_{5}$. The original ASD simulation data are shown for comparison. The additional curves are the results of the ASD simulation where the change of the absolute value of Dy moment is taken into account. Two sets of exchange couplings J$_{ij}$ are used, calculated for 47 K and for 78.9 K.} 
\label{Kuzmin}
\end{figure}

One factor not accounted for in the simulations is the temperature dependence of the absolute values of the magnetic moments. The ASD method considers only the thermal distribution of the moment orientations but does not include variations in the magnitudes of the moments with temperature. To estimate the influence of this effect, we adopted the following procedure. In the DMFT+HIA approach, the Fermi smearing parameter corresponds directly to the bare electronic temperature. Using this relation, we can reproduce the temperature dependence of the Dy magnetic moment below 100 K as described in \cite{HERPER2023118473} for NdFe$_{11}$Ti. The calculated temperature evolution of Dy spin, orbital, and total moments is shown in Fig.~\ref{Dymom}, along with the experimental data from \cite{PhysRevB.96.024412}. The total Dy moment values were then fitted using the Kuz'min formula (Eq.~\ref{eq:kuzmin}) and used as input in the ASD simulation, thereby allowing the absolute value of the Dy moments to vary with temperature. Since the Co moments are not expected to change significantly, their magnitudes were kept fixed.The resulting behavior is shown in Fig.~\ref{Kuzmin}. We find that this simplified approximation -- neglecting the change in Co moments -- leads to a substantial increase in the total magnetization at higher temperatures. Although this result overshoots the experimental values, it suggests that the discrepancy between the experimental data and the ASD simulation results is likely due to the neglect of the temperature dependence of the magnetic moment magnitudes.

Another factor commonly neglected in ASD simulations is the temperature dependence of the exchange couplings $J_{ij}$. To assess its effect on the magnetization curve, we performed an additional simulation using J$_{ij}$s values computed with DMFT+HIA at a higher temperature of 78.9 K. As shown in Fig.~\ref{Kuzmin}, this adjustment results in an increased Curie temperature, which likely contributes to the remaining mismatch between the original ASD simulations and the experimental results.

\section{The origin of the anisotropy of magnetization} \label{sec:app2}

DMFT+HIA calculations performed for two orientations of the spin axis -- namely,in-plane and out-of-plane -- yield Dy 4{\it f} magnetic moments that differ by nearly 1.5 $\mu_B$. We investigated the origin of this discrepancy in detail. In our approach, the double-counting correction $\hat{{\bf H}}^{DC}_R=\mu_{DC}\hat{{\bf 1}}+\hat{{\bf H}}_X$ is applied to the Dy 4{\it f} states, where $\mu_{DC}$ denotes the atomic chemical potential and $\hat{{\bf H}}_X$ is constructed to remove the local intra-orbital LDA exchange splitting (for details, see \cite{GRANAS2012295}). The term $\hat{{\bf H}}_X$ can be divided into intra-orbital $\hat{{\bf H}}_X$ and inter-orbital contributions $\hat{{\bf H}}_{XX}$. The latter term is obtained in a separate self-consistent calculation in which the 4{\it f} electrons are constrained to remain paramagnetic, and is subsequently subtracted from the local Hamiltonian of the spin-polarised calculation \cite{GRANAS2012295}. By analysing the individual contributions to $\hat{{\bf H}}_{XX}$, we identified the terms that contribute the most to the 4{\it f} magnetic moment: the chemical potential, the spin-orbit coupling, the chemical potential, and the charge quadrupole moment. Among these, the charge quadrupole term was found to be the dominant factor responsible for the change in the magnetic moment upon rotation of the spin axis. The charge quadrupole moment is defines as $w^{2020}=3[l(2l-1)]^{-1}Q_{ZZ}$, where $Q_{ZZ}=\sum_i(l^2_z-\frac{1}{3}l^2)_i$, as detailed in \cite{PhysRevB.80.035121}. The dominant contribution of the quadrupole moment causes re-ordering and re-distribution of the ground state energies in DyCo$_{5}$, see  Table~\ref{table:GS}.

The importance of the quadrupole term for the rare-earth elements was noted in several works on RE-based materials \cite{PhysRev.132.706,PhysRevB.39.12548}.

\begin{table*}[h!]
\centering
\caption{Ground state information for the in-plane and out-of-plane orientation of spin axes. We list only the levels with non-negligible weight ($w$). The energy is given in mRy. Where any of $J_z,L_z,S_z,S_x,S_y,Q_{zz}$ are not listed, they are equal to zero. For all levels, $J=7.50$, $L=5.10$, $S=2.43$}
\begin{tabular}{c c } 
 \hline \hline
in-plane &  out-of plane\\ 
 \hline
$E=0$ $w=0.86$, $S_x=2.41$, $Q_{zz}=-2.53$ & $E=0$ $w=0.29$, $J_z=6.50$, $L_z=4.39$, $S_z=2.11$, $Q_{zz}=2.83$  \\ 
$E=0.6$ $w=0.11$, $S_x=2.08$, $Q_{zz}=-1.87$ & $E=0.01$ $w=0.28$, $J_z=7.50$, $L_z=5.09$, $S_z=2.41$, $Q_{zz}=4.84$  \\ 
& $E=0.06$ $w=0.24$, $J_z=5.50$, $L_z=3.69$, $S_z=1.81$, $Q_{zz}=1.13$  \\ 
& $E=0.2$ $w=0.13$, $J_z=4.50$, $L_z=3.00$, $S_z=1.50$, $Q_{zz}=-0.28$  \\ 
\hline \hline
\end{tabular}
\label{table:GS}
\end{table*}

\section{Previous work} \label{sec:app3}


Here, we present a brief overview of the investigations conducted on DyCo$_5$ up to the present. Significant experimental and theoretical work has been done to understand the magnetic properties of the material. Various theoretical models have been proposed, and different methods tested, to reproduce experimental findings such as the spin reorientation and compensation points. However, a fully satisfactory description has yet to be found, partly due to the lack of high-quality experimental data measured on single crystals. All magnetic data are summarized in Table~\ref{table:overview}.

In \cite{PhysRevB.96.024412}, the authors present both experimental measurements of the magnetic moment (including orbital and spin contributions) and theoretical simulations. The authors use the experimental crystal structure of DyCo$_5$ for their calculations, performed within the framework of Korringa-Kohn-Rostoker (KKR) method, using the Atomic Sphere Approximation (ASA) and the Local Spin-Density Approximation (LSDA). The 4{\it f} electrons of Dy were treated within the frozen-core approximation, assuming a spin moment of 5 $\mu_{\bf B}$ and an orbital moment of 5 $\mu_{\bf B}$ according to the Hunds rules. The {\it spd} moment was calculated to be -0.50 $\mu_{\bf B}$. Self-consistent (SC) calculations were performed for both ferrimagnetic (FI) and ferromagnetic (FM) states, with the former being lower in energy by 1.76 mRy/atom. The Co moment was calculated to be $m_{Co(I)}$ = -1.58 $\mu_{\bf B}$ and $m_{Co(II)}$ = -1.50 $\mu_{\bf B}$, resulting in a total moment of 1.76 $\mu_{\bf B}$ per formula unit (f.u.).

Isotropic exchange interactions were obtained using the magnetic force theorem for atomic pairs within the distance of 2{\it a} for the atomic pairs. The nearest neighbours (NN) Co-Co exchange interactions were found to be approximately 1.1 mRy, where Dy and the Co(II) moments were coupled antiferromagnetically with an NN interaction of -0.23 mRy. 

A spin Hamiltonian, including the Heisenberg exchange constants, uniaxial anisotropy constant, and basal plane anisotropy constant, was used for simulations based on the stochastic Landau-Lifshitz-Gilbert (LLG) equation of motion. The magnetic anisotropy energy (MAE) values were determined with LDA{\it + U} ($U=7$ and $J=0.7$ eV) within
the KKR formalism. These values were -1.4 mRy and 0.17 mRy for Dy, and 0.0083 and 0.064 mRy for the Co sublattice. The MAE for Co was adjusted to 0.1 mRy per ion for each Co sublattice to reach the reorientation transition.  As a result of the simulations, the CT was found to be $T_{comp} = 120$ K, while spin reorientation occurred between T$_{\bf SR1}$ of 320 K and T$_{\bf SR2}$ of 360 K.

Magnetizations, CT, and SR temperatures were also measured in DyCo$_5$ thin films \cite{PhysRevB.96.024412}. The X-ray magnetic circular dichroism (XMCD) technique was employed in the temperature range from 10 K to 475 K to measure changes in magnetic anisotropy and the variation of the elemental magnetic moments. The sum-rules \cite{VOGEL1993170} equations were used to determine the orbital and spin moments of both Co and Dy. At the lowest temperature, the spin magnetic moment of Co was measured to be 1.62 $\mu_{\bf B}$/atom, with an orbital moment of 0.13 $\mu_{\bf B}$/atom. For Dy, the measured spin and orbital moments were 4.47 $\mu_{\bf B}$/atom and 4.97 $\mu_{\bf B}$/atom, respectively, when extrapolated to 0 K.

ASD simulations (similar to those discussed in  \cite{PhysRevB.96.024412}) and experimental investigations (XMCD), were performed in \cite{Damping}, where the authors explored the difference in damping parameters between Dy and Co. According to the XMCD data, Co demagnetizes within approximately 200fs, while  Dy takes about 1ps. By minimizing the mean squared error between the XMCD data and the ASD simulation results, the authors obtained element-speciﬁc damping parameters of $\alpha_{Co}=0.004$ and $\alpha_{Dy}=0.03$.

Full-potential LSDA and LSDA{\it + U} were used in \cite{MILETIC2007604} to calculate the magnetic moment of DyCo$_5$, with the results summarized in Table~\ref{table:overview}. The magnetization direction was chosen to lie in the plane perpendicular to the {\it c}-axis, in accordance with experimental data. Although the size of the magnetic moment in the LSDA calculation was close to experimental values, ferromagnetic ordering was incorrectly predicted to be more stable than ferrimagnetic ordering, with an energy difference of 18.5 mRy. The LSDA{\it + U} method was used to correct the issue, with $U_{4f}$ set to 0.4 Ry.

A DFT-DLM approach was introduced in \cite{PhysRevB.97.224415}, which accounts for finite-temperature effects in the disordered local moment (DLM) approximation \cite{PhysRevB.74.144411,Gyorffy_1985}. In this approach, the 4{\it f} electrons were treated within DFT, using KKR in combination with the coherent potential approximation (CPA), ASA, and SIC. However, this method tends to overestimate the magnetic moment of Dy, while underestimating the moment of the Co sublattice.

In \cite{SMARDZ1999209}, X-ray photoelectron spectroscopy (XPS) data were compared to the density of states (DOS). The DOS calculations were performed using the tight-binding linear muffin-tin orbital (TB LMTO) method in the ASA approximation, using experimental lattice parameters. The authors employed the scalar-relativistic approximation for the band electrons, while the core electrons were treated relativistically as a frozen core. The 4{\it f} states of Dy were treated as open-core states.

Another XMCD investigation was conducted in \cite{PhysRevB.104.054439} on DyCo$_{4.6}$ thin ﬁlms. 

Several experimental investigations report the values of magnetic moment, as well as CT and SR temperatures \cite{TSUSHIMA1983197,DyCo5exp,Schweizer_1980}, the numbers  are listed in Table~\ref{table:overview}. \cite{PhysRev.140.A131} presents M{\"o}ssbauer measurements of the magnetic moment,  macroscopic magnetization measurements are reported in \cite{10.1063/1.1728809}. 

In \cite{Kelaeev}, the intersublattice exchange parameter and the microscopic constant of the second-order anisotropy (per rare-earth ion) were derived from experimental data based on the phenomenological theory of magnetocrystalline anisotropy. These values are $I_{Dy-Co}=0.5$ mRy, $K_{Dy}=0.04$ mRy.

Although the magnetism of DyCo$_5$ has been discussed in numerous publications, relatively few data on magnetocrystalline anisotropy energy (MAE) have been presented. An overview of the MAE and crystal field parameters is provided in Table~\ref{table:ani}. The MAE for the {\it f} ions were derived from the single-ion approximation in  \cite{RADWANSKI1986120,GREEDAN1973387}. Neutron diffraction data were interpreted in terms of the phenomenological theory of magneto-crystalline anisotropy in \cite{Kelaeev}.

\section{Residual moment at compensation point} \label{sec:app4}

\begin{figure}[h]
 \centering
 \includegraphics[scale=0.5]{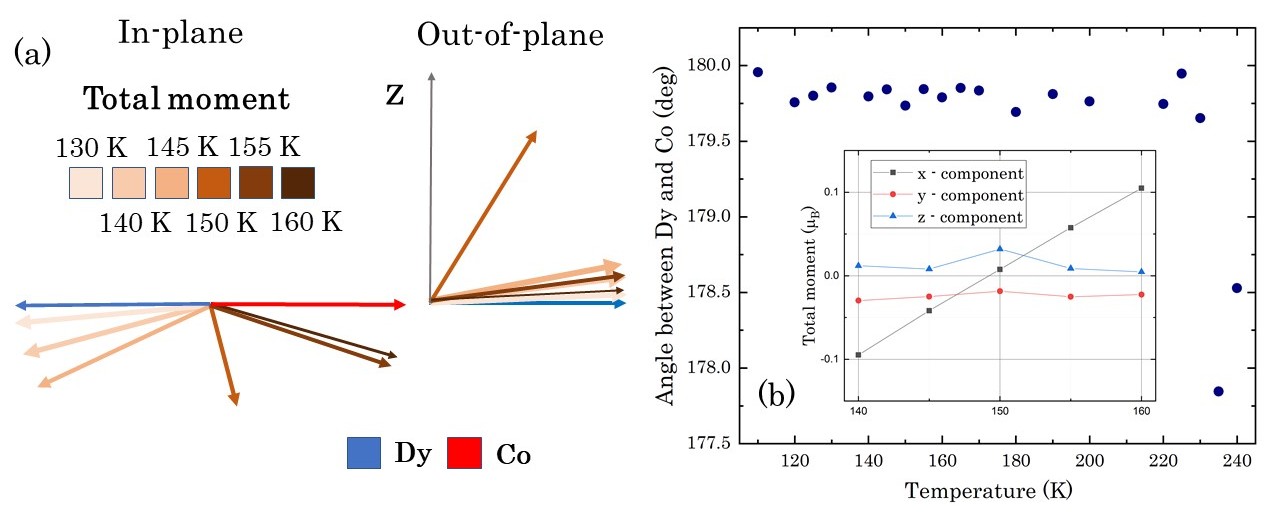}
\caption{(Color online) (a) Rotation of the in-plane and out-of-plane components of the total moment with temperature. Note that the lengths of the Dy, Co, and total moments are normalised for better visibility. (b) The change in the angle between the Dy and Co moments with temperature. (inset) The x, y, and z components of the total moment, where spin-quantization axis is {\it x}.} 
\label{Compensation}
\end{figure}

To investigate the origin of the residual moment at the compensation point,we examined in detail the orientation of the total moment as well as those of the Dy and Co sublatices. A small angular deviation (tilt) is observed between M$_{Dy}$ and M$_{Co}$ within the {\it ab}-plane. The tilt is non-zero even at low temperatures and increases progressively with temperature (see Figure~\ref{Compensation} {\it (b)}).  Figure~\ref{Compensation} {\it (a)} shows the evolution of the total-moment direction near the compensation point. As the total Co and Dy sublattice moments become equal along the quantisation axis {\it (x)}, see the inset of Figure~\ref{Compensation} {\it (b)}, a slight tilt causes the total moment to rotate within the plane rather than vanish completely. We attribute the origin of the subtle in-plane Dy-Co sublattice misalignment to intrinsic non-collinearity, driven by the exchange interactions, as the strong anisotropy term forces the moments to remain mostly in-plane up to the reorientation temperature. To confirm that this effect is not an artifact of finite-size limitations in our ASD simulations, we extended the supercell size to $35\times35\times35$, finding that the tilt persisted.

Figures~\ref{Compensation} {\it (a)} and {\it (b)} also reveal a small out-of-plane component that increases slightly around the compensation point. This feature is most likely attributable to thermal disorder. Around the compensation temperature, the Dy sublattice already exhibits a very high angular spread, see Fig.~\ref{M_T_polar} of the main text, which significantly impacts the precise cancellation of magnetization vectors. There is a progressive increase in tilt around the compensation point, where sublattice magnetizations are most delicately balanced. Note, that the moments in Figure~\ref{Compensation} {\it (a)} are normalised; the residual moment is too small to be shown to scale.

Although the residual moment is negligible in the absence of an external field, it increases significantly when a magnetic field is applied in the ASD simulations. The applied field enhances the in-plane tilt between the sublattices, and together with the thermal spread of the Dy moments, produces a pronounced nonzero total magnetization -- consistent with the experimental observations under high-field conditions.

\section{Orbital contribution to the exchange interactions} \label{sec:app5}

\begin{figure}[h]
 \centering
 \includegraphics[scale=0.5]{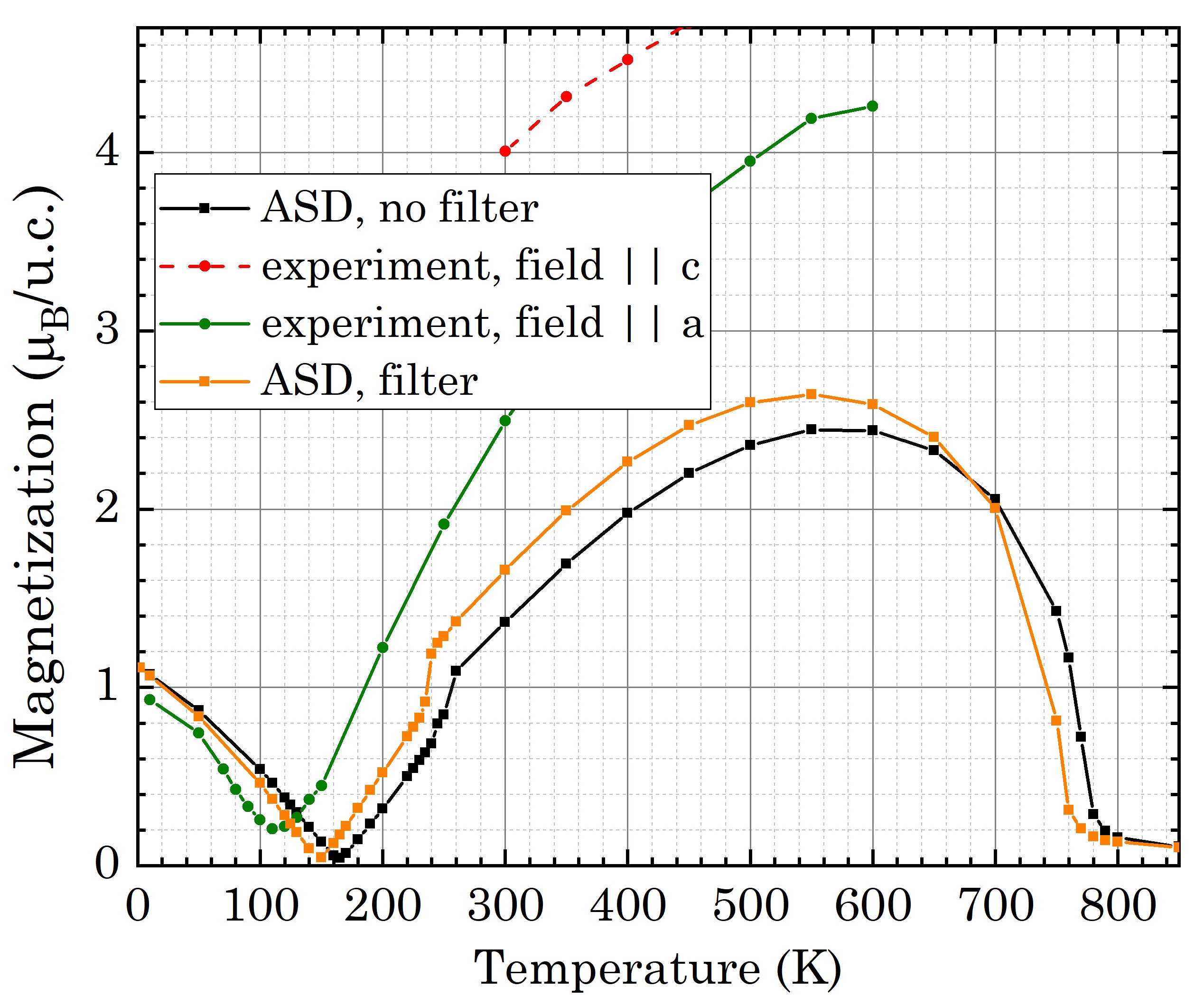}
\caption{(Color online) The change in the M(T) curve for the J$_{ij}$s calculated with and without filtering compared to experimental data.} 
\label{filter}
\end{figure}

The implementation of the exchange interaction calculation in the relativistic framework within RSPt is described in detail in \cite{PhysRevB.102.115162}. In this approach, the system is perturbed by rotating the initial magnetic moment by a small angle. As is shown in the {\it Methods} section, the final expression for $J_{\alpha \beta}$ contains commutators between the Hamiltonian and spin operators. When SO coupling is included in the Hamiltonian, these commutators in equation~(\ref{eq:HASD}) induce orbital rotation that can modify the chemical bonding. To prevent this, only the spin exchange terms are retained in the commutator, while the other terms are filtered out. Here, we illustrate the impact of this filtering on the magnetization curves, see Figure~\ref{filter}. As expected, modifying the exchange parameters influences the thermal distribution of the magnetic moments, thereby affecting both the compensation and Curie temperatures. However, the magnitude of this change is modest and significantly smaller than the effects discussed in {\it  Appendix}~\ref{sec:app1}.

\end{document}